\documentclass[a4paper,11pt]{article}	
\pdfoutput=1 

\newcommand{\ket}[1]{\left| #1 \right\rangle}

\newcommand{\grad}{\vec\nabla}

\usepackage[table]{xcolor}                  
\usepackage{colortbl}
\usepackage{booktabs}                       
\usepackage{multirow,bigdelim}              
\usepackage{units}                          
\usepackage{amsmath,amssymb,mathtools}      
\usepackage{amsthm}
\usepackage{bbm,dsfont}                     
\usepackage{ytableau}                       
\usepackage{empheq}                         
\usepackage{graphicx}                       
\usepackage{subfig}                         
\usepackage{float}
\usepackage[numbers,sort&compress]{natbib}  
\usepackage[colorlinks=true,urlcolor=blue,anchorcolor=blue,citecolor=blue,filecolor=blue,linkcolor=blue,menucolor=blue,pagecolor=blue,linktocpage=true,pdfproducer=medialab,pdfa=true]{hyperref}	      
\usepackage{verbatim}

\newlength{\xtrawidth}
\newlength{\xtraheight}
\numberwithin{equation}{section}           
\numberwithin{table}{section}
\numberwithin{figure}{section}
\renewcommand{\labelenumi}{\emph{(\roman{enumi})}}

\graphicspath{{figs/}}

\makeindex                                 

\def\md{\mathbf}

\def\ms{\mathsf}

\def\bF{\mathbb{F}}

\def\bP{\mathbb{P}}
\def\bQ{\mathbb{Q}}
\def\bR{\mathbb{R}}
\def\bZ{\mathbb{Z}}

\def\cA{\mathcal{A}}

\def\cD{\mathcal{D}}

\def\cH{\mathcal{H}}

\def\cN{\mathcal{N}}
\def\cO{\mathcal{O}}
\def\cP{\mathcal{P}}

\def\sx{\mathsf{x}}
\def\sy{\mathsf{y}}

\def\sH{\mathsf{H}}
\def\sO{\mathsf{O}}
\def\stO{\tilde{\mathsf{O}}}

\def\sQ{\mathsf{Q}}

\def\Tx{\mathsf{T}_{\sx}}
\def\Ty{\mathsf{T}_{\sy}}
\def\TTx{\tilde{\mathsf{T}}_{\sx}}
\def\TTy{\tilde{\mathsf{T}}_{\sy}}
\def\xa{x_\alpha}
\def\xb{x_\beta}
\def\ya{y_\alpha}
\def\yb{y_\beta}
\def\dex{\delta x}
\def\dey{\delta y}
\def\dpx{\delta x^\prime}
\def\dpy{\delta y^\prime}

\def\ket#1{\left| #1 \right\rangle}

\DeclarePairedDelimiterX\opm[3]{\langle}{\rangle}{#1 \delimsize\vert #2 \delimsize\vert #3}
\DeclarePairedDelimiterX\ip[2]{\langle}{\rangle}{#1 \delimsize\vert #2}

\def\({\left(}
\def\){\right)}
\def\[{\left[}
\def\]{\right]}

\newcommand{\re}{{\rm e}}
\newcommand{\ri}{{\mathsf{i}}}
\newcommand{\rd}{{\rm d}}

\newcommand{\pd}{\partial}

\newcommand{\nn}{\nonumber \\}

\newcommand{\be}{\begin{equation}}
\newcommand{\ee}{\end{equation}}
\newcommand{\ba}{\begin{aligned}}
\newcommand{\ea}{\end{aligned}}

\def\Catalan{\mathrm{C}}

\title{\boldmath We shall need one}
\author{Zhihao Duan, Jie Gu, Yasuyuki Hatsuda, Tin Sulejmanpasic}

\begin{document}
\begin{titlepage}
	{}~ \hfill\vbox{ \hbox{} }\break

	\rightline{LPTENS 18/10}
	\rightline{RUP-18-18}
	
	\vskip 1 cm

	\begin{center}
		\Large \bf  Instantons in the Hofstadter butterfly: difference equation, resurgence and quantum mirror curves
		\end{center}
	
	\vskip 0.8 cm
	
	\centerline{Zhihao Duan$^\dagger $, Jie Gu$^\dagger$, Yasuyuki Hatsuda$^\sharp$, Tin Sulejmanpasic$^\star$}
	
	\vskip 0.2in	
	\begin{center}{\footnotesize
			\begin{tabular}{c}
				{\em $^\dagger$Laboratoire de Physique Th\'eorique \& $^\star$Institut de Physique Th\'eorique Philippe Meyer,}\\
				{\em \'Ecole Normale Sup\'erieure, CNRS, PSL Research University, Sorbonne Universit\'es, UPMC} \\
				{\em 24 rue Lhomond, 75231 Paris Cedex 05, France} \\[2ex]
				{\em $^\sharp$Department of Physics, Rikkyo University} \\
				{\em Toshima, Tokyo 171-8501, Japan} \\[2ex]
				
				{\em $^\star$ Institute for Nuclear Physics, University of Mainz, D-55099 Mainz, Germany}
			\end{tabular}
		}
	\end{center}

	\setcounter{footnote}{0}
	\renewcommand{\thefootnote}{\arabic{footnote}}
	\vskip 60pt
	\begin{abstract}
		
	We study the Harper-Hofstadter Hamiltonian and its corresponding non-perturbative butterfly spectrum. The problem is algebraically solvable whenever the magnetic flux is a rational multiple of $2\pi$. For such values of the magnetic flux, the theory allows a formulation with two Bloch or $\theta$-angles. We treat the problem by the path integral formulation, and show that the spectrum receives instanton corrections. Instantons as well as their one loop fluctuation determinants are found explicitly and the finding is matched with the numerical band width of the butterfly spectrum. We extend the analysis to all 2-instanton sectors with different $\theta$-angle dependence to leading order and show consistency with numerics. We further argue that the instanton--anti-instanton contributions are ambiguous and cancel the ambiguity of the perturbation series, as they should. 
		We hint at the possibility of exact 2-instanton solutions responsible for such contributions via Picard-Lefschetz theory. We also present a powerful way to compute the perturbative fluctuations around the 1-instanton saddle as well as the instanton--anti-instanton ambiguity by using the topological string formulation.
	\end{abstract}
		
	{
		\let\thefootnote\relax
		\footnotetext{Emails: duan@lpt.ens.fr, jie.gu@lpt.ens.fr, yhatsuda@rikkyo.ac.jp, tin.sulejmanpasic@gmail.com}
	}
		
	\end{titlepage}
	\vfill \eject

	\newpage
	\tableofcontents
	
\vspace{4ex}

\section{Introduction}

The spectral problem for electrons on a two-dimensional square lattice in a uniform magnetic field was originally considered by Harper in 1955 \cite{Harper:1955}, where an elegant difference equation was derived. More than 20 years later in  1976 Hofstadter derived a recursive equation which allowed him to plot the spectrum as a function of the magnetic field, now known as the Hofstadter butterfly \cite{Hofstadter:1976zz}. Due to the magnetic effect, the electron spectrum shows a rich structure.
Recently, a novel link between a two-dimensional electron lattice system and a Calabi-Yau geometry was found in \cite{Hatsuda:2016mdw}.
It was pointed out in \cite{Hatsuda:2016mdw} that this Hofstadter's spectral problem is related to another spectral problem appearing in the mirror geometry of the toric Calabi-Yau manifold known as local $\bF_0$ \cite{Grassi:2014zfa}.  The interesting point of this relation is that the magnetic effect is interpreted as a kind of \textit{quantum deformations} of the Calabi-Yau geometry.  
One can probe quantum Calabi-Yau geometry by the 2d electron lattice system in the magnetic field.
The correspondence was generalized to the triangular lattice and another Calabi-Yau manifold \cite{Hatsuda:2017zwn}.

In the present paper, our goal is a more quantitative understanding of this relation as well as the non-perturbative and resurgent structure of the spectrum. We here focus on the band structure of the Harper-Hofstadter problem in the weak magnetic limit. In this regime,  we can treat the magnetic flux perturbatively. The perturbative expansion of the energy spectrum can explain the position (the center) of the band for each Landau level.
However, it does not explain the width of bands because the band width is \textit{non-perturbative} in the weak magnetic flux limit.
Such non-perturbative corrections are caused by quantum mechanical tunneling effects. 
We will demonstrate that the non-perturbative band width is explained by instanton effects in the path integral formalism. This was observed long ago in \cite{PhysRevB.41.11328} (see also \cite{Wilkinson305} for the WKB approach to the problem). However here we will focus on the resurgent properties intimately related to these instantons, or more correctly to the multi-instanton contributions which we discuss in some details.

Technically, we have a very efficient way to compute the perturbative expansion of the energy spectrum around the trivial saddle \cite{Sulejmanpasic:2016fwr, Gu:2017ppx}, but this efficient way is not applicable for the computation of semiclassical expansion around the other nontrivial saddles.
To our knowledge, there are no systematic ways to compute the semiclassical expansions around the instanton saddles in the Harper-Hofstadter model. We employ several approaches to extract this information. One is a brute force numerical approach, which we use as a check. The second is a path-integral approach, where we find the exact saddle of the path-integral action and the one-loop fluctuation. We push this computation to the 2-instanton sector and find matching results to the numerics. The instanton analysis is performed only to the leading-order in perturbation theory, and is not easily extended to perturbative corrections around the instanton saddles.

To extract corrections around instanton saddles, we employ a rather unconventional approach. We use the connection with a toric Calabi-Yau threefold, local $\bF_0$, and find that the non-perturbative band width is captured by the free energy of the refined topological string on this geometry. Using this remarkable connection, we can efficiently compute the semiclassical fluctuation around the 1-instanton saddle by using the string theory technique, called the refined holomorphic anomaly equations \cite{Bershadsky:1993cx, Krefl:2010fm, Huang:2010kf}. 
Our approach here is conceptually very similar to the previous works \cite{Codesido:2017jwp} on certain quantum mechanical systems\footnote{In fact, the results in \cite{Codesido:2017jwp} correspond to the special case, the midpoint of each sub-band in our analysis.}. We would like to emphasize that here we have a realistic electron system where string theory techniques can be applied.

The structure of the rest of the paper is as follows. In section~\ref{sc:H-model} we quickly review the eigenvalue problem of the Harper-Hofstadter model and its exact solutions when the magnetic flux $\phi$ is $2\pi$ times a rational number. We argue that in the latter case there are two Bloch's angles which can be turned on, while only one of them can be turned on if the magnetic flux has a generic value, as in the case of a trans-series solution of the Harper-Hofstadter model. In section~\ref{sc:1-inst}, we make a trans-series ansatz for the energy in the small $\phi$ limit. We then compute the leading order contribution in the 1-instanton sector for the ground state energy by a path integral calculation, and find that it agrees with the numerical results.
In section~\ref{sc:2-inst}, we perform further path integral calculations in the 2-instanton sector, and compare with the numerical results. The imaginary part of the instanton--anti-instanton sector is extracted numerically using the well-known relation to the large order growth of the perturbative energy. Inspired by \cite{Codesido:2016dld, Codesido:2017jwp, Fischbach:2018yiu}, we also find in section~\ref{sc:flucturation} the fluctuations in the 1-instanton and instanton--anti-instanton sector can be computed from topological string on local $\bF_0$. 
Finally we conclude and list some open problems in section~\ref{sc:conclusion}.

\section{The Harper-Hofstadter problem}
\label{sc:H-model}

To prepare for the other sections, we quickly review in this section the classic results on the Harper-Hofstadter model \cite{Harper:1955,Hofstadter:1976zz}, including the formulation of its eigenvalue problem, and the exact solutions when the magnetic flux is $2\pi$ times a rational number. We make the careful distinction that there are two Bloch's angles in this case while only one of them can be turned on if the value of the magnetic flux is generic.

\subsection{The eigenvalue problem of the Harper-Hofstadter equation}
\label{sc:Hamiltonian}

The Harper-Hofstadter model describes an electron in a two dimensional lattice potential with a uniform magnetic flux in the perpendicular direction. Let the lattice spacing be $a$, and suppose the electron momentum has components $k_x$ and $k_y$ in the two directions. The energy of the electron before turning on the magnetic flux is, up to a normalization
\begin{equation}
	E = -\frac{1}{2}(\re^{\ri k_x a} + \re^{-\ri k_x a} + \re^{\ri k_y a} + \re^{-\ri k_y a}) +2\ .
\end{equation}
We have chosen for later convenience a particular normalization so that the energy vanishes for zero electron momentum. 
In this convention, the energy forms a single band $0\leq E \leq 4$.

After we turn on the magnetic flux, quantum mechanically we get the Hamiltonian operator by replacing the momentum $\vec{k}$ by the operator\footnote{We work in $\hbar=c=1$ units.} $\vec{\pi}:=\vec{p} - \vec{A}$. Notice that $\vec p$ is the \emph{canonical momentum}. Upon the gauge transformation $\vec A\rightarrow \vec A+\grad \Lambda$, the Hamiltonian is only invariant up to a canonical transformation $\vec p\rightarrow \vec p+\grad\Lambda$. Under such a canonical transformation, the state of the Hilbert space transforms as $\ket{\Psi}\rightarrow e^{i\Lambda (x, y)}\ket{\Psi}$.
Notice that the momentum $\vec \pi$ generally depends on the coordinates. Indeed this is reflected in the fact that the commutator
\be
	[\pi_x,\pi_y]=\ri F_{xy}(x,y)
\ee
where $F_{xy}(x,y)=\pd_x A_y-\pd_y A_x$ is the $xy$ component of the field-strength tensor of $\vec A$, i.e.\ the magnetic field through the $xy$-plane at the point $(x,y)$. Henceforth, we consider the case where the magnetic field is uniform: $F_{xy}(x,y)=B$. 

Replacing\footnote{Despite the notation, $\sx$ and $\sy$ are not the original coordinates of the system, but are proportional to the magnetic translation operators.} $\sx=\pi_xa, \sy=\pi_ya$, we have that the lattice Hamiltonian becomes
\be\label{eq:Hof_Hamiltonian}
	H= -\frac{1}{2}(\re^{\ri \sx } + \re^{-\ri \sx} + \re^{\ri \sy } + \re^{-\ri \sy}) +2\ .
\ee
with the commutation relation
\be\label{eq:xyphi}
	[\sx,\sy]=\ri\phi,
\ee
where $\phi=B a^2$ is the flux of the magnetic field through the plaquette. We will also use the exponentiated notation
\begin{equation}\label{eq:Txy}
	\Tx = \re^{\ri \sx}\ ,\quad \Ty = \re^{-\ri \sy}
\end{equation}
with the commutation relation
\begin{equation}\label{eq:TxTy-com}
	\Tx \Ty = \re^{\ri \phi} \Ty \Tx
\end{equation}
so that the Hamiltonian can be written as
\begin{equation}\label{eq:Hof_THamiltonian}
	H = -\frac{1}{2}(\Tx + \Tx^{-1} + \Ty + \Ty^{-1})+2 \ .
\end{equation}

We regard $\sx$ and $\sy$ as the canonical operators, and can now look at eigenstates $\ket\psi$ in the $\sx$-representation, i.e.\ define $\psi(x)=\left\langle x |\psi\right\rangle$ where $\ket{x}$ is an eigenstate of $\sx$ with eigenvalue $x$, so that
\be\label{eq:eigeneq}
	H\ket\psi=E\ket\psi\Rightarrow -\frac{1}{2}\left(\psi(x+\phi)+\psi(x-\phi)\right)-\cos(x)\psi(x)=(E-2)\psi(x)
\ee
which is just the difference equation.

\subsection{Symmetries and $\theta$-angles}

The Hamiltonian \eqref{eq:Hof_Hamiltonian} clearly commutes with the symmetry operators%
\footnote{Similar operators also play an important role in the context of quantum mechanics associated with toric Calabi-Yau threefolds \cite{Hatsuda:2016mdw}.}
\be
	\TTy=\re^{\ri \frac{2\pi \sx}{\phi}}\;, \TTx=\re^{-\ri\frac{2\pi \sy}{\phi}}\;,
\ee
each of which generates a group $\mathbb Z$. The labelling above is because 
\be
	\TTy \;\sy\;\TTy^\dagger=\sy - 2\pi\;,\qquad \TTx\;\sx\;\TTx^\dagger=\sx - 2\pi\ .
\ee
 But we generally have
\be
	\TTx\TTy=\re^{-\ri \frac{4\pi^2}{\phi}}\TTy\TTx \ .
\ee
Since for a generic value of $\phi\in \mathbb R$ the operators commute up to a phase, we can say that the physical symmetry group $\mathbb Z\times \mathbb Z$ acts projectively. 

Let us first choose that
\begin{equation}\label{eq:overQ}
	\phi=2\pi/Q \ ,\quad Q\in\mathbb Z \ .
\end{equation}
In that case the two operators commute, and the symmetry $\mathbb Z\times \mathbb Z$ is no longer acting projectively. Now we can project to simultaneous eigenstates of the operators $\TTx$ and $\TTy$, i.e.\ we can demand that
\be
	\TTx\ket\Psi=\re^{\ri\theta_x}\ket\Psi\;, \TTy\ket\Psi=\re^{\ri\theta_y}\ket\Psi
\ee
The angles $\theta_x$ and $\theta_y$ are Bloch's angles for the $x$ and $y$ translations. Notice however that they can only be defined in this way if $2\pi/\phi\in \mathbb Z$. 

Next, we consider more general case that
\be\label{eq:rationality}
	\phi/(2\pi) = P/Q \in \mathbb Q \ ,
\ee
where $P,Q$ are coprime integers. 
Then we have that
\be
	\TTx\TTy=\re^{-\ri\frac{2\pi Q}{P}}\TTy\TTx\;.
\ee
Clearly, if $P\neq 1$, the generators $\TTx,\TTy$ must be supplemented by the generator\footnote{$\ms I_P$ is equivalent to $\re^{-\frac{2\pi \ri Q}{P}}$ because there always exists an integer $k$ such that $\re^{-\frac{2\pi \ri Q}{P}k}=\ms I_P$} $\ms I_P=\re^{\ri\frac{2\pi}{P}}$, and the $\mathbb Z\times\mathbb Z$ must be centrally extended by $\mathbb Z_P$.

What about $\theta$-angles? In this case we have that $[(\TTx)^{P},(\TTy)^{P}]=0$, and we can define $\theta_x,\theta_y$ angles by the simultaneous eigenstate of $(\TTx)^{P}$ and $(\TTy)^{P}$. Alternatively, in this case we also have $[(\TTx),(\TTy)^{P}]=0$, so we could equally define the two $\theta$-angles as eigenstates of these two operators. Finally if $P=n^2$ is a perfect square, we have that $[(\TTx)^{n},(\TTy)^{n}]=0$ and we can define $\theta$-angels accordingly as well. In most cases we will only consider $P=1$. Then, all these definitions of $\theta$-angles coincide, and we are back to the scenario \eqref{eq:overQ}.

Finally if $\phi/2\pi$ is irrational, then 
\be
	\TTx\TTy=\re^{\ri\alpha}\TTy\TTx \ ,
\ee
where $\alpha/(2\pi) = -2\pi/\phi$ is irrational as well.
The additional generator $\ms I_\alpha=\re^{\ri\alpha}$ generates the group $\mathbb Z$, so the $\mathbb Z\times \mathbb Z$ is centrally extended by $\mathbb Z$. In this case we are allowed only one $\theta$-angle, which we can get as an eigenstate of either $\TTx$ or $\TTy$ but not both simultaneously.

\subsection{Exact solutions for rational magnetic flux}
\label{sc:exact}

It is well-known that the eigenvalue problem \eqref{eq:eigeneq} can be solved exactly if the rationality condition \eqref{eq:rationality} is satisfied \cite{Hofstadter:1976zz}. 
Let us set
\begin{equation}\label{eq:phi-ab}
	\phi = 2\pi P/Q
\end{equation}
where $P,Q$ are two coprime integers and $Q>0$. 
The underlying reason of the exact solvability is that in the case of \eqref{eq:phi-ab} we can project onto simultaneous eigenstates of the powers $\TTx^P$ and $\TTy^P$, as these two operators commute. This will allow, as we shall see, for a finite-dimensional representation of the operators $\Tx$ and $\Ty$, in which the Hamiltonian \eqref{eq:Hof_THamiltonian} is written, and give us an algebraic equation for the eigenvalue problem.
Note that in this case $\Tx,\Ty$ are also shift operators, as
\begin{equation}
	\Tx \,\sy\, \Tx^{\dagger} = \ms \sy -2\pi P/Q\ ,\quad \Ty \,\sx\, \Ty^{\dagger} = \ms \sx - 2\pi P/Q \ .
\end{equation}

Recall that in this case we can define $\theta$-angles as eigenvalues of $\TTx^P=\Ty^Q$ and $\TTy^P=\Tx^Q$.
Now let us for the moment choose $\theta_x=\theta_y=0 \bmod 2\pi$, i.e.
\be
	(\Tx^{(0)})^Q=(\Ty^{(0)})^Q=\mathbf{1}\;,
\ee
In other words we impose periodic boundary conditions on physical states under the shift $\sx\rightarrow\sx-2\pi P$ and $\sy\rightarrow\sy-2\pi P$. The algebra \eqref{eq:TxTy-com}, which now reads
\be\label{eq:rat_alg}
	\Tx^{(0)}\Ty^{(0)}=\re^{\frac{2\pi \ri P}{Q}}\Ty^{(0)}\Tx^{(0)}
\ee
has a finite dimensional representation in terms  of the clock and shift matrices
\begin{equation}
	\Tx^{(0)} = \begin{pmatrix}
	1 & 0 & 0 & \ldots & 0 \\
	0 & q & 0 & \ldots & 0\\
	0 & 0 & q^2 & \ldots & 0 \\
	\vdots & \vdots & \vdots &  & \vdots \\
	0 & 0 & 0 &\ldots & q^{b-1} 
	\end{pmatrix} \ , \quad 
	\Ty^{(0)} = \begin{pmatrix}
	0 & 0 & \ldots & 0& 1\\
	1 & 0 & \ldots & 0 & 0 \\
	0 & 1 & \ldots & 0 & 0 \\
	\vdots & \vdots & & \vdots & \vdots \\
	0 & 0 & \ldots & 1 & 0
\end{pmatrix} \ ,
\end{equation}
where  $q = \re^{\ri\phi} = \re^{\frac{2\pi\ri P}{Q}}$. Note also that $(\Tx^{(0)})^Q=(\Ty^{(0)})^Q=\mathbb I_{Q\times Q}$, as it should.

Now let us introduce the twisted boundary condition through the replacement $(\Tx^{(0)},\Ty^{(0)})\rightarrow (\Tx,\Ty)=(\Tx^{(0)} \re^{\ri\frac{\theta_x}{Q}},\Ty^{(0)} \re^{\ri\frac{\theta_y}{Q}})$. Then we have that
\be
	(\Tx)^Q=\re^{\ri\theta_x} \mathbf{1}, \quad (\Ty)^Q=\re^{\ri\theta_y} \mathbf{1}\;,
\ee
while the algebra \eqref{eq:TxTy-com} is intact. Alternatively, the twisted boundary condition is equivalent to a deformation of the Hamiltonian. Using the notation $k_x = \theta_x/Q, k_y = \theta_y/Q$, we can write the Hamiltonian operator depending on $k_x$ and $k_y$ as
\be
	H(k_x, k_y)= -\frac{1}{2}(e^{\ri k_x}\Tx^{(0)} + e^{-\ri k_x} \Tx^{(0)-1} + e^{\ri k_y}\Ty^{(0)} +e^{-\ri k_y} \Ty^{(0)-1})+2,
\label{eq:H(kx,ky)}
\ee
while keeping the boundary condition periodic. Now we are finally ready to write the eigenvalue equation for the operator \eqref{eq:Hof_Hamiltonian}. 
Plugging the matrix representation of $\Tx^{(0)}$ and $\Ty^{(0)}$ into \eqref{eq:H(kx,ky)}, 
the Hamiltonian becomes
\be
	H(k_x,k_y)=\begin{bmatrix}
	2-\cos(k_x) & -\tfrac{1}{2}\re^{-\ri k_y} & 0 & \ldots & 0 &   -\tfrac{1}{2}\re^{\ri k_y} \\
	-\tfrac{1}{2}\re^{\ri k_y} & 2-\cos\left(k_x+\frac{2\pi P}{Q}\right)& -\tfrac{1}{2}\re^{-\ri k_y} & \ldots & 0 &  0 \\
	\vdots & \vdots & \vdots & & \vdots & \vdots  \\
	0 & 0 & 0 & \ldots &  & -\tfrac{1}{2}\re^{-\ri k_y} \\
	-\tfrac{1}{2}\re^{-\ri k_y} & 0 & 0 & \ldots & -\tfrac{1}{2}\re^{\ri k_y} & 2-\cos\left(k_x+\frac{2\pi(Q-1)P}{Q}\right)
	\end{bmatrix}
\ee
so that the characteristic equation $\det(H-E\, \mathbb I_{Q\times Q})=0$ is given by

\begin{equation}\label{eq:Fab}
	F_{P/Q}(E,k_x,k_y) = \det \begin{bmatrix}
	M_0 & -\re^{-\ri k_y} & 0 & \ldots & 0 &  0 & -\re^{\ri k_y} \\
	-\re^{\ri k_y} & M_1 & -\re^{-\ri k_y} & \ldots & 0 & 0 & 0 \\
	\vdots & \vdots & \vdots & & \vdots & \vdots & \vdots \\
	0 & 0 & 0 & \ldots & -\re^{\ri k_x} & M_{Q-2} & -\re^{-\ri k_y} \\
	-\re^{-\ri k_y} & 0 & 0 & \ldots & 0 & -\re^{\ri k_y} & M_{Q-1}
	\end{bmatrix} = 0 \ ,
\end{equation}
with
\begin{equation}
	M_n = 2(2-E) - 2\cos(2\pi n P/Q + k_x) \ .
\end{equation}
As in \cite{Hasegawa:1990}, it is straightforward to check that
\begin{equation}
	F_{P/Q}(E,k_x,k_y) = F_{P/Q}(E,k_x,0) - 2\cos(Q k_y) +2 \ .
\end{equation}
Using the symmetry under the mapping $(k_x,k_y) \mapsto (k_y,-k_x,)$, one finds that the equation \eqref{eq:Fab} can be simplified to
\begin{equation}\label{eq:Fab-simp}
F_{P/Q}(E,0,0) + 4 = 2(\cos(\theta_x) + \cos (\theta_y)) \ ,
\end{equation}
with the Bloch's angels $\theta_x = Qk_x, \theta_y = Q k_y$.
It is then a simple job to get eigen-energy $E$ by solving \eqref{eq:Fab-simp}.

\begin{figure}
	\centering
	\includegraphics[width=0.5\linewidth]{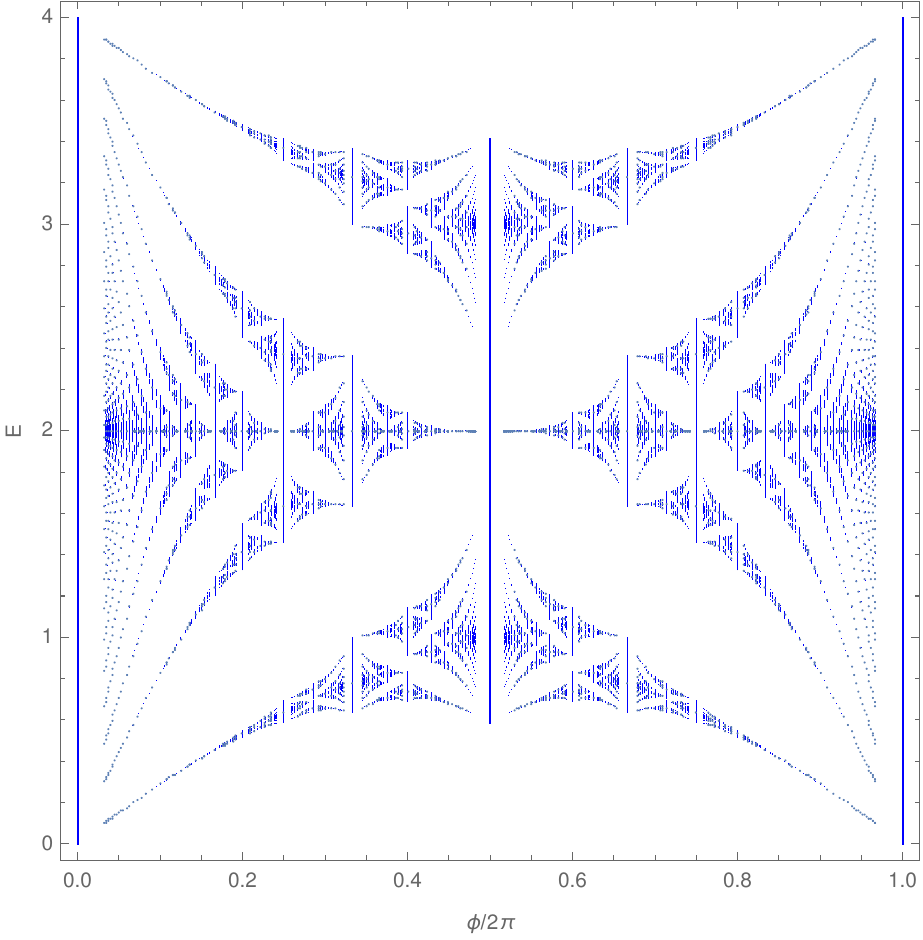}
	\caption{The Hofstadter butterfly plots energy levels $E_{k_x,k_y}(N,\phi)$ with $0\leq k_x, k_y \leq 2\pi/Q$ against magnetic flux $\phi \in 2\pi \bQ$ for the Harper-Hofstadter model. We take $\phi/2\pi$ to be $P/Q$ for any coprime pairs of positive integers such that $P\leq Q$ and $Q\leq 30$.}\label{fg:butterfly}
\end{figure}

We notice that the equation \eqref{eq:Fab-simp} depends on the value of $P$ only through the polynomial $F_{P/Q}(E,0,0)$. Note \eqref{eq:Fab-simp} indicates that the minimal ranges for the Bloch's angles $\theta_x,\theta_y$ are
\begin{equation}
	0 \leq \theta_x < 2\pi \ ,\quad 0 \leq \theta_y < 2\pi \ ,
\end{equation}
as they should.
By varying the values of $\theta_x,\theta_y$, the eigen-energies $E(\theta_x, \theta_y)$ form bands. The two edges of a energy band correspond to 
$(\theta_x,\theta_y) = (0,0), (\pi,\pi)$. If we turn off one Bloch's angle, the energy band width is reduced to its one half. We reproduce in Fig.~\ref{fg:butterfly} the famous plot of the Hofstadter butterfly, which is a plot of the energy bands as a function of the magnetic flux $\phi$ when $\phi/2\pi$ is rational.

\section{Trans-series expansion and 1-instanton sector}
\label{sc:1-inst}

\subsection{Why trans-series expansion?}

We are interested in the energy spectrum of the Harper-Hofstadter model in the weak flux limit $\phi\to 0$. As discussed in section~\ref{sc:H-model}, with generic values of $\phi$, we should use the Hamiltonian operator \eqref{eq:Hof_Hamiltonian} for the twisted boundary condition with only one Bloch's angle.
Throughout this paper, however, we consider the weak flux limit with the specific form
\be
\phi=\frac{2\pi}{Q},\qquad Q \to \infty,
\label{eq:weak-phi}
\ee
then we can introduce two distinct Bloch's angles $\theta_x$ and $\theta_y$ simultaneously.

We want to understand the spectral behavior in the limit \eqref{eq:weak-phi}.
To do so, it is useful to treat $\phi$ as a continuous parameter even in the specific case \eqref{eq:weak-phi}. 
Since the Hamiltonian is a Laurent polynomial of $\re^{\ri \sx}$ and $\re^{\ri \sy}$, we can use the \texttt{Mathematica} package \texttt{BenderWu} \cite{Sulejmanpasic:2016fwr, Gu:2017ppx} to compute its perturbative energy\footnote{The Hamiltonians considered in \cite{Gu:2017ppx} consist of operators $\re^{\sx}$ and $\re^{\sy}$ with $[\sx, \sy]=\ri \hbar$. To translate it into our case here, one has to identify $\hbar=-\phi$. See subsection~\ref{sc:flucturation} for detail. We have however updated the \texttt{BenderWu} package version 2.2 with a function \texttt{BWDifferenceArray}, which allows of mixed inclusion of terms $e^{x},e^{p},e^{\ri x},e^{\ri p}$. The package is available on Wolfram Package site.}.
The first few orders are as follows
\begin{equation}\label{eq:E-pert}
\begin{aligned}
	E^\text{pert}(N) 
	=\frac{2N+1}{2}\phi-\frac{2N^2+2N+1}{16}\phi^2+\frac{2N^3+3N^2+3N+1}{384}\phi^3
	+\cO(\phi^4) \ ,
\end{aligned}
\end{equation}
where $N$ is the Landau level of the eigen-energy. We note the agreement with earlier studies \cite{0305-4470-24-10-019, 10.21468/SciPostPhys.4.5.024}.

The perturbative energy \eqref{eq:E-pert} or even its Borel resummation cannot be the full answer. First of all, the higher order terms of the perturbative series have the same sign, and thus its Borel transform of the perturbative series has poles on the positive axis, leading to ambiguity in the Borel resummation. This ambiguity is an indication that the energy receives non-perturbative corrections. We discuss the ambiguity in detail in section~\ref{sc:growth}.  Second, the perturbative series clearly does not depend on Bloch's angles, thus itself alone cannot explain the energy bands. 
As a result, the band spectrum should have the trans-series expansion, with the explicit dependence of $\theta_x$ and $\theta_y$ in instanton sectors.
The trans-series expansion of the spectrum should take the following form:
\begin{equation}\label{eq:E-transseries}
\begin{aligned}
	E_{(\theta_x, \theta_y)}(N)=E^\text{pert}(N)+E_{(\theta_x, \theta_y)}^\text{1-inst}(N)+E_{(\theta_x, \theta_y)}^\text{2-inst}(N)+\cdots\quad (\phi \to 0)\ . \\
\end{aligned}
\end{equation}
The leading perturbative contribution is given by \eqref{eq:E-pert}.
The $k$-instanton sector is exponentially suppressed by a factor $\re^{-kA/\phi}$ with a constant $A$.%
\footnote{More precisely, beyond the one-instanton order, in general logarithmic corrections of the form $\log^\ell \phi$ also appear. Therefore, the full trans-series expansion consists of three kinds of trans-monomials: $\phi$, $\re^{-A/\phi}$ and $\log \phi$.}

Our goal in this paper is to reveal this trans-series structure.
In particular, we will show explicit forms for a few instanton sectors.
In the subsequent subsections, we will first compute the leading (1-loop) order contribution to the 1-instanton for the ground state energy by an honest path integral computation, and then
compare them with the prediction from the numerical analysis.
We further argue 
that the quantum fluctuations in the one-instanton sector for any energy level 
can be read off from the topological string theory on local $\bF_0$.
In the next section, we will investigate the 2-instanton sector.

\subsection{Path integral in one-instanton sector}
\label{sc:1-instanton}

 The problem of instantons in the Harper/Hofstadter problem was first discussed in \cite{PhysRevB.41.11328} where the authors computed the one-instanton and its one-loop determinant numerically. Here we will re-derive these instanton solutions and compute analytically the one-loop fluctuations in the instanton sectors of the ground state energy. 

 To begin with, let's reproduce the Hamiltonian operator \eqref{eq:Hof_Hamiltonian} for convenience
\begin{equation}\label{eq:H-norm}
	\sH(\sx,\sy):= -\cos \sx -\cos \sy + 2 \ ,\qquad
	[\sx, \sy]=\ri \phi\, .
\end{equation}
The cosine potential has infinitely many degenerate vacua located at
\begin{equation}
	x = 2\pi n_x \ ,\;  y=2\pi n_y \ ,\quad n_x,n_y\in \bZ \ .
\end{equation}
Classically we have complete freedom of whether to identify different vacua as physically equivalent. This is not possible quantum mechanically for generic values of $\phi$ as we shall see.

Treating $\phi$ as the Planck constant, the above Hamiltonian can be associated with the Euclidean path integral
\begin{equation}\label{eq:Zbeta}
	Z =\text{tr }\re^{-\frac{\beta}{\phi}\sH(\sx,\sy)}= \int \cD x \cD y \exp\[-\frac{1}{\phi}\int_{-\beta/2}^{\beta/2} \rd t \(H(x,y) - \ri \dot{x} y \) \] \ .
\end{equation}
with boundary conditions for $x$ and $y$ to be specified momentarily. The partition function above is related to the eigen-energies $E(N)$ with levels $N=0,1,2,\ldots$ of the Hamiltonian $H$ by
\begin{equation}
	Z = \sum_{N=0}^\infty \re^{-\beta E(N)/\phi} \ ,
\end{equation}
so that the ground state energy $E(0)$ can be obtained through the Euclidean path integral in the large $\beta$ limit.

Before we continue we should emphasize that the action of the above path integral is similar to that of the phase-space quantum mechanical system where $x$ is identified with a coordinate, and $y$ is identified with a momentum. The difference is that here we do not have a purely Gaussian dependence on the ``momentum'' $y$. For this reason we cannot integrate it out. Still one may hope to analyze the problem semi-classically. But there are several issues here. Firstly the semi-classics of path-integrals is to this day not a completely understood subject, but it has become clear recently that the correct interpretation of it is via the Picard-Lefschetz (PL) theory \cite{Witten:2010zr,Witten:2010cx,Harlow:2011ny,Behtash:2015kna,Behtash:2015kva,Behtash:2015zha,Behtash:2015loa,Kozcaz:2016wvy,Behtash:2018voa,Nekrasov:2018pqq}. The PL theory analysis is by far not a straightforward matter, and requires the identification of saddles which contribute in the semi-classical expansion. As we shall see all such saddles of the action above will be on complex $x,y$ trajectories. We do not a priori know whether such saddles should contribute. To determine it we should compute the so-called intersection number of the co-thimble (we refer the reader to the cited literature for details). This is a difficult task way beyond our current understanding. We will find some instanton solutions and argue that they must contribute on physical grounds. We will check quantitatively their contribution against numerics and find exact agreement. 

Secondly it is not clear whether a continuum limit of the above path-integral exists. The path-integral is typically obtained by slicing the Boltzmann weight into $N$ pieces, and inserting a complete set of states in between. This amounts to a lattice discretization of the path-integral, with a lattice spacing $\epsilon=\beta/N$. Upon integration over the momentum, the resulting path-integral has a Gaussian suppression factors $\re^{-(\dots)\frac{(x_{i+1}-x_i)^2}{\epsilon}}$. As we take the continuum limit $\epsilon\rightarrow 0$ the path of $x$ is forced to be smoother and smoother. No such smoothness seems to be justified in the continuum-limit of the phase-space path integral above. Still as we shall see the semiclassical analysis passes many non-trivial checks against the numerical brute-force calculation.

The boundary conditions of the path integral can be made strictly periodic. This amounts to saying that values of coordinates $(x,y)$ and $(x+2\pi n_x,y+2\pi n_y)$ are physically distinct for any $n_x,n_y\in \mathbb Z$. In this case the above Lagrangian has a shift symmetry which takes $x\rightarrow x+2\pi n_x$ and $y\rightarrow y+2\pi n_y$, with $n_{x,y}\in \mathbb Z$. 

Now let us consider the values of $x$ and $x+2\pi$ to be physically equivalent. In other words we are gauging the shift symmetry of the scenario above, projecting the full Hilbert space down to eigenstates of a shift symmetry operator. Without a $\theta_x$-term, the projection will be to singlets of the shift operator. Gauging the symmetry amounts to saying that the boundary conditions must be relaxed to include periodicity of $x(t)$ up to a $2\pi$ shift, i.e.\ $x(t+\beta)=x(t)+2\pi m_x$, where $m_x$ is to be summed over. The integers $m_x$ can be viewed as holonomies of the $\mathbb Z$-valued gauge field which we have to sum over in order to project to a subspace of singlets under the shift symmetry $x\rightarrow x+2\pi$.

Notice however that after gauging the $x$-shift symmetry, shifting $y$ to $y+2\pi n_y$ we get an additional phase in the partition function
\be
	\re^{\frac{\ri}{\phi}(2\pi)^2 n_y m_x}\;.
\ee
The above is only unity if $\phi=2\pi/Q$, where $Q\in \mathbb Z$. Hence if we insist that $x\sim x+2\pi$  (i.e.\ $x$-shift symmetry is gauged) and that $y\rightarrow y+2\pi$ is a global symmetry we must have that\footnote{From the point of view of the Hilbert space this means that if $x\rightarrow x+2\pi$ is a gauge symmetry, the operator which shifts $\sy\rightarrow \sy+2\pi$, given by $\re^{\ri 2\pi \sx/\phi}$, is not a gauge invariant operator unless $2\pi/\phi\in \mathbb Z$, and even though it commutes with the Hamiltonian, it is not a valid generator of the symmetry transformation.}  $\phi=2\pi/Q$. This is of course evident from the Hilbert space picture, but it is satisfying to see it in the path-integral. Incidentally we can say that there is a 't~Hooft anomaly between the two $(\mathbb Z)_x$ and $(\mathbb Z)_y$ shift symmetries, so that the system must break at least one of the two to saturate the anomaly.

Since we are assuming that $\phi=2\pi/Q$, we can insert the two $\theta$-angles by introducing the terms $\theta_y \frac{\dot y}{2\pi}$ and $\theta_x\frac{\dot x}{2\pi}$.
The path integral can be treated by the saddle-point approximation if $\phi$ is small. The main contribution comes from the perturbative saddle for which $x=y=0$ at any time $t$. This solution does not break the translational symmetry on the time-circle, and all its modes are Gaussian. The perturbative partition function can be expanded in powers of $\phi$ using the Feynman diagrams. The result will be the perturbative partition function which we denote as $Z_0$. In turn this is related to the perturbative energies as follows
\be
	Z_0=\sum_{N=0}^\infty \re^{-\beta E^{\text{pert}}(N)/\phi}\;.
\ee
where $E^{\text{pert}}(N)$ is the perturbative energy at level $N$.

On the other hand, the contributions of the partition function can be classified by their topological winding number, i.e.\
\begin{equation}
	Z(\beta,\theta) = Z_0 + Z_1+Z_{-1} +Z_2+Z_{-2} \cdots = Z_0 \(1+ \sum_{n\ne 0}^\infty \hat{Z}_n\) \ ,\; \; \hat{Z}_n = Z_n/Z_0 \ ,
\end{equation}
where $Z_0$ is the expansion around the trivial saddle point (i.e.\ the perturbative vacuum), and it is responsible for perturbative contributions $E^\text{pert}(0)$
\begin{equation}
	Z_0 \approx C\re^{-\beta E^\text{pert}(0)/\phi} \ ,\qquad \beta \to \infty\ ,
\end{equation}
while $Z_{n\ne 0}$ come from different instanton sectors ($n$ counts the instanton number). The constant $C$ above may be UV divergent, and may be removed by the appropriate definition of the path integral measure. Further all $Z_n$-s are UV divergent. However all the UV divergences are the same, and so $\hat Z_n$ is UV finite. The constant $C$ therefore factorizes, and is of no physical consequence as it cancels in the observables.

The \emph{dilute instanton gas approximation} makes now the following assumption: the multi-instanton contributions factorize to 1-instanton contributions.  So 
\be
	\hat Z_n=\sum_{m-\bar m=n}\frac{\hat{Z}_1^{m}}{m!}\frac{\hat{Z}_{-1}^{\bar m}}{\bar m!}\ .
\ee
Summing over $n$ we simply have
\be\label{eq:dilute}
	Z(\beta,\theta)\approx Z_{\text{dilute instanton gas}}=Z_0 \,\re^{\hat{Z}_1+\hat{Z}_{-1}} \ .
\ee
Now the $\hat{Z}_{\pm 1}$ is given by
\be\label{eq:Zhat1}
	\hat{Z}_{\pm 1} = -\int_{-\beta/2}^{\beta/2} \rd t\; K \re^{-A/\phi\, \pm \ri\theta}= - \beta K \re^{-A/\phi\,\pm \ri\theta}
\ee
where $K$ is the measure of the $1$-instanton configurations, including the perturbative corrections, and $\theta$ is the relevant $\theta$-angle coupling to the instantons\footnote{In the Harper-Hofstadter problem we will have two types of instantons which tunnel in $x$- and $y$-directions respectively. So we may have two $\theta$-angles: $\theta=\theta_x$ or $\theta=\theta_y$ coupling to the tunneling events $x\rightarrow x+2\pi$ and $y\rightarrow y+2\pi$. Recall that these $\theta$ angles can only be defined when $2\pi/\phi\in \mathbb Z$.}.
Therefore the 1-instanton correction to the ground state energy is given by
\begin{equation}\label{eq:E1loop}
	E_\theta^\text{1-inst}(0)=E_{I}+E_{\bar I} = 2\phi K \re^{-A/\phi}\cos\theta\ .
\end{equation}
To get this correction we need to compute $K$.

Let us first consider the partition function $Z(\beta,\theta)$ in the trivial vacuum given by $x=y=0$ by expanding in $x$ and $y$ up to quadratic terms and performing the Gaussian integral to get
\begin{equation}
	Z_0(\beta) \approx \frac{1}{(\det \sO_0)^{1/2}}\ ,
\end{equation}
with
\begin{equation}\label{eq:O0}
	\sO_0 = -\partial^2_t + 1 \ ,
\end{equation}

\begin{figure}
	\centering
	\subfloat[$x$-profile]{\includegraphics[width=0.4\linewidth]{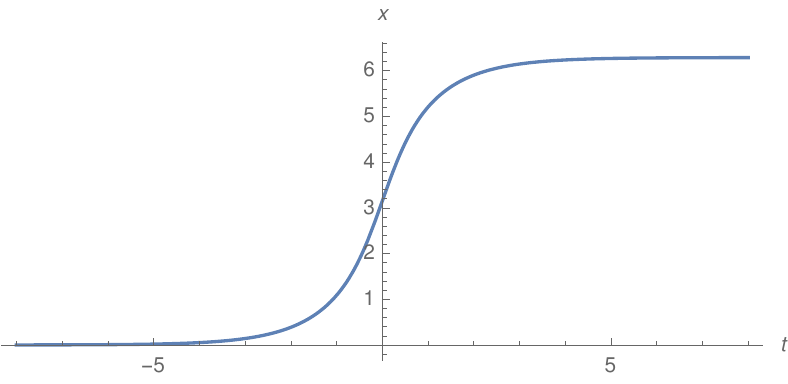}}\hspace{2ex}
	\subfloat[$y$-profile]{\includegraphics[width=0.4\linewidth]{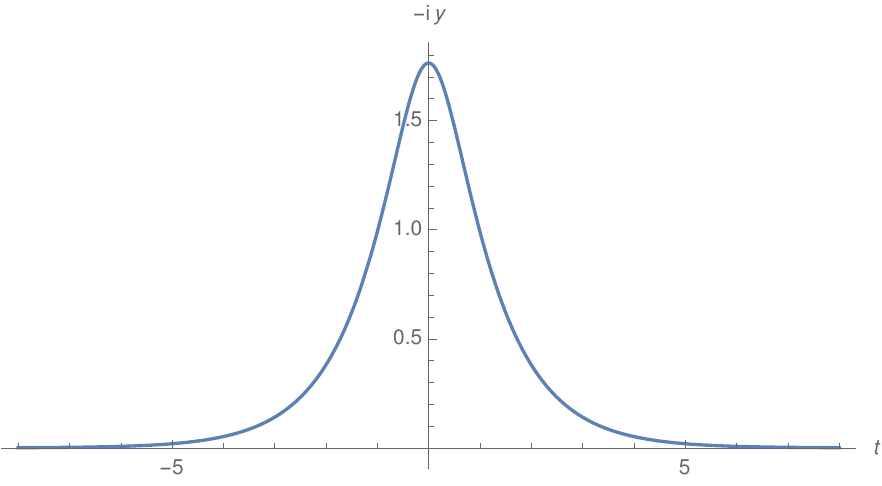}}
	\caption{The $x$- and $y$-profiles of 1-instanton in the Harper-Hofstadter model. The value of $y$ is purely imaginary.}\label{fg:inst-profile}
\end{figure}

Now we consider the 1-instanton sector. For this purpose, we need to solve for the 1-instanton configuration. The equations of motion for the partition function \eqref{eq:Zbeta} is
\begin{subequations}\label{eq:eom}
	\begin{align}
	& \ri \dot{x} - \sin y = 0 \ , \label{eq:eom1}\\
	& \ri \dot{y} + \sin x = 0 \ . \label{eq:eom2}
	\end{align}
\end{subequations}
We solve these equations in the Appendix \ref{app:1-instanton} to give the 1-instanton solution
\begin{align}\label{eq:inst1}
	&x_1(t) = 2\cos^{-1} \( -\frac{\sqrt{2}\tanh(t-t_0)}{\sqrt{1+\tanh^2(t-t_0)}}\) \ ,
	&y_1(t) = \cos^{-1} \( 1+\frac{2}{\cosh2(t-t_0)}\) \ ,
\end{align}
where $t_0$ is a free parameter interpreted as the center of the instanton. Note that $x_1(t)$ starts from 0 in $t= -\infty$ and reaches $2\pi$ in $t = +\infty$, and thus it indeed has topological charge 1, while $y_1(t)$ is always imaginary and its imaginary value reaches the maximum $\cos^{-1}(3)$ at $t=t_0$. 
This means that we are considering the instanton tunneling in the $x$-direction.
We call it an $x$-instanton.
We plot $x_1(t)$ and $-\ri y_1(t)$ in Fig.~\ref{fg:inst-profile}. 
The profile of an anti-instanton is obtained by simply the time-reversal transformation\footnote{The time reversal transformation takes $T: (x(t),y(t))\rightarrow (x(-t),-y(-t))$. In addition we have a parity transformation which takes $P: (x(t),y(t))\rightarrow (-x(t),-y(t))$. }. We also notice that the Hamiltonian is constant
\begin{equation}\label{eq:inst1-Esimp}
	\cos y_1 + \cos x_1 = 2\ ,
\end{equation}
as it should be,
with the help of the e.o.m.\ \eqref{eq:eom1}, which will be of use later.

There exists in fact another type of 1-instanton due to the fact that the Hamiltonian function is also periodic in $y$. In the example of the Harper-Hofstadter model, one can easily find the new instanton due to the symmetry of the theory under the map 
\begin{equation}
	(x(t),y(t)) \to (-y(t),x(t)) \ .
\end{equation}
Applying this map to the instanton solution \eqref{eq:inst1}, we get a new instanton solution with the $x$- and $y$-profiles exchanged (up to a minus sign). We call it a $y$-instanton, since it has a non-trivial topological charge in the $y$-direction, but a trivial topological charge in the $x$-direction. This instanton does not couple to $\theta_x$. Instead it couples to the $\theta_y$-angle.\footnote{We remind the reader that both $\theta_x$ and $\theta_y$ are only possible if the $2\pi/\phi\in\mathbb Z$, which we assume here. However much of the results will hold for generic $\phi$, as we shall comment later.
}

Let us compute the action of the 1-instanton configuration \eqref{eq:inst1}, in the limit $\beta \rightarrow \infty$. The action of the instanton is computed 
analytically in appendix~\ref{app:1-instanton} and it reads
\be
	A=8\Catalan\ ,
	\label{eq:A-inst}
\ee
where $\Catalan$ is the Catalan's constant. 

Now we compute the one-loop partition function in the 1-instanton sector, by performing the expansion
\begin{equation}
	x = x_1 + \delta x \ ,\quad y = y_1 + \delta y \ ,
\end{equation}
and keeping only terms up to quadratic orders. Using the conservation law \eqref{eq:inst1-Esimp} as well as the e.o.m.\ \eqref{eq:eom}, we have
\begin{equation}
	Z_1(\beta) \approx {\re^{-A/\phi+\ri\,\theta}} \int\cD (\delta x) \cD (\delta y) \exp\[ -\frac{1}{2\phi}\int_{-\beta/2}^{\beta/2} \rd t \(\cos x_1 \cdot \delta x^2 + \cos y_1 \cdot \delta y^2 -2\ri \,\delta\dot{x} \delta{y} \) \] \ .
\end{equation}
We can first integrate out $\delta y$. However notice that in doing so we will get a nontrivial factor in front of the path-integral, because the coefficient of $\delta y^2$ is not a constant. To avoid this, let us first replace $\delta \tilde y=\sqrt{\cos y_1}\delta y$ and $\delta \tilde x=\delta x/\sqrt{\cos y_1}$.\footnote{Note that $\cos y_1>0$, because $y_1$ is purely imaginary on the instanton trajectory.} Notice that this replacement keeps the measure invariant i.e.\ $\cD(\delta \tilde x)\cD(\delta\tilde y)=\cD(\delta  x)\cD(\delta y)$. Upon integrating out the $\delta \tilde y$, we get
\begin{align}\label{eq:inst_inter}
	\nonumber Z_1(\beta) \approx & {\re^{-A/\phi+\ri\theta}}  \\ &\times \int\cD (\delta \tilde x) \exp\[ -\frac{1}{2\phi}\int_{-\beta/2}^{\beta/2} \rd t \({ \frac{\left[\partial_t\left({\delta \tilde x\sqrt{\cos y_1}}\right)\right]^2}{\cos y_1}}+\cos x_1\cos y_1 \delta \tilde x^2 \) \] \\\nonumber
	=&\frac{{\re^{-A/\phi+\ri\,\theta}}}{\sqrt{\det\stO}}
\end{align}
where the operator $\stO$ is
\be\label{eq:Otilda}
	\stO=-{\sqrt{\cos y_1}}\partial_t \frac{1}{\cos y_1}\; \partial_t {\sqrt{\cos y_1}}+\cos y_1\cos x_1\;.
\ee
The operator $\stO$ has a zero mode given by $\psi_0(t)=N^{-1}\frac{\dot x_1(t)}{\sqrt{\cos y_1}}$, as can be checked. Here 
\be\label{eq:zeromode_norm}
	N=\sqrt{(\dot x_1/\sqrt{\cos y_1},\dot x_1/\sqrt{\cos y_1})}
\ee 
is the normalization factor.
So the above expression of the one loop weight of the instanton cannot be correct. The zero mode originates from the time-translation symmetry of the theory. In other words, field fluctuations which only change the location of the instanton do not change the action, and the modes in this direction must be treated exactly (i.e.\ beyond the Gaussian approximation).

To find the measure of the instanton we must first separate out the zero mode, which we denote by $t_0$. We will get that
\be
	Z_1=\re^{-A/\phi+\ri\theta}\int \rd t_0\;  \; \frac{\mu}{\sqrt{\det' \stO}}\ ,
\ee
where the prime indicates that the zero mode has been excluded from the determinant. The $\mu$ above is the measure of the instanton moduli $t_0$ (or the moduli space metric). It is given by (see Appendix \ref{app:moduli-space metric})
\be
\mu=\sqrt{\frac{N^2}{2\pi \phi}}\ ,
\ee
so that the one-loop instanton contribution to the partition function is given by
\begin{equation}
	Z_1(\beta) \approx \int\frac{\rd t_0}{\sqrt{2\pi\phi}} \sqrt{(\dot{x}_1/\sqrt{\cos y_1(t)},\dot{x}_1/\sqrt{\cos y_1(t)})} \frac{\re^{-A/\phi+\ri\,\theta}}{(\det' \sO)^{1/2}} \;.
\end{equation} 
The contribution is of course divergent, as the functional determinant is infinite in the continuum.
We therefore normalize it with respect to the perturbative partition function. The normalized 1-instanton partition function is given by
\begin{equation}\label{eq:hatZ1}
	\hat{Z}_1(\beta) = \frac{Z_1(\beta)}{Z_0(\beta)} \approx \re^{-A/\phi+\ri\,\theta}\frac{\beta\sqrt{(\dot{x}_1/\sqrt{\cos y_1},\dot{x}_1/\sqrt{\cos y_1})}}{\sqrt{2\pi\phi}}  \(\frac{\det \sO_0}{\det' \stO}\)^{1/2} \ .
\end{equation}
Comparing with \eqref{eq:Zhat1}, we find that the prefactor $K$ entering formula \eqref{eq:E1loop} is given by\footnote{A possible minus sign can be absorbed into the $\theta$ angle}
\be\label{eq:K_res}
	K=\frac{\sqrt{(\dot{x}_1/\sqrt{\cos y_1},\dot{x}_1/\sqrt{\cos y_1})}}{\sqrt{2\pi\phi}}  \(\frac{\det \sO_0}{\det' \stO}\)^{1/2}\;.
\ee
As we show in the Appendix \ref{app:determinant}, the ratio of determinants is given by
\begin{equation}
	\frac{\det\nolimits'\stO}{\det \sO_0} = \frac{\dot{x}_1(-\beta/2)\dot{x}_1(\beta/2)}{\sinh \beta\cos y_1(-\beta/2)}\int_{-\beta/2}^{\beta/2}\rd t\, \frac{\dot{x}^2_1(t)}{\cos y_1(t)}\int_{-\beta/2}^{t}\rd t' \frac{\cos y_1(t')}{\dot{x}^2_1(t')}\int^{\beta/2}_{t}\rd t'' \frac{\cos y_1(t'')}{\dot{x}^2_1(t'')} \ .
\end{equation}
Note that in obtaining the above result, we have used Dirichlet boundary conditions for the space of function acted on by the operators $\sO$ and $\sO_0$. Since we will only be interested in the limit $\beta\rightarrow \infty$, the boundary conditions will not matter. However if one is interested in computing the instanton contributions to higher energy levels, a computation with periodic boundary conditions is necessary.

Further since we only care about the limit $\beta\rightarrow \infty$, we can make convenient approximations. We notice that the 1-instanton configuration \eqref{eq:inst1-x}, \eqref{eq:inst1-p} has the following asymptotic form
\begin{equation}\label{eq:inst-asym}
	\dot{x}_1(t) \sim A_\pm \re^{\mp\omega t} \ ,\;\cos y_1(t) \sim 1+B_\pm \re^{\mp 2\omega t} \ , \quad t\rightarrow \pm \infty \ ,
\end{equation}
where
\begin{equation}
	A_\pm = 2\sqrt{2} \ ,\; B_\pm = 4 \ ,\; \omega = 1 \ .
\end{equation}
Besides, the integrand of the integral over $t$ is small when $t$ is close to $\pm \beta/2$, so the integral over $t$ is saturated away from them.
So regarding the two integrals over $t'$ and $t''$, only the $-\frac{\beta}{2} \ll t\ll \frac{\beta}{2}$ region is important.  The two integrals can be approximated by
\begin{equation}
\begin{aligned}
	&\int_{-\beta/2}^{t}\rd t' \frac{\cos y_1(t')}{\dot{x}^2_1(t')} \sim \int_{-\beta/2}^{t}\rd t' \frac{\re^{-2\omega t'}}{A_-^2} \sim \frac{\re^{\omega \beta}}{2\omega A_-^2} \ ,\\
	&\int_{t}^{\beta/2}\rd t'' \frac{\cos y_1(t'')}{\dot{x}^2_1(t'')} \sim \int_{t}^{\beta/2}\rd t'' \frac{\re^{2\omega t''}}{A_+^2} \sim \frac{\re^{\omega \beta}}{2\omega A_+^2} \ .
\end{aligned}
\end{equation}
Pulling these two integrals out of the integral of $t$, the latter becomes $(\dot{x}_1/\sqrt{\cos y_1},\dot{x}_1/\sqrt{\cos y_1})$. Apply \eqref{eq:inst-asym} in the remaining part of the determinant evaluation, we find in the end
\begin{equation}
	\frac{\det\nolimits'\stO}{\det \sO_0} = \frac{(\dot{x}_1/\sqrt{\cos y_1},\dot{x}_1/\sqrt{\cos y_1})}{16} \ .
\end{equation}
we get that the factor $K$ in \eqref{eq:K_res} is given by
\be\label{eq:K}
	K= 4\(\frac{1}{2\pi\phi}\)^{1/2}.
\ee
The anti-instanton partition function is the same but with the opposite topological charge. Therefore using \eqref{eq:E1loop}, the leading order 1-instanton correction to the ground state energy given by $x$-instanton coupled to $\theta_x$ is
\begin{equation}
	E_{I_x}(0) + E_{\bar{I}_x}(0) = 8 \cos \theta_x \(\frac{\phi}{2\pi}\)^{1/2}\re^{-A/\phi} \ .
\end{equation}
Since we have two kinds of instantons coupled to $\theta_x$ and $\theta_y$ respectively, the full 1-instanton correction is finally given by
\be
\ba
	E_{(\theta_x,\theta_y)}^\text{1-inst}(0)&=E_{I_x}(0)+E_{\bar I_x}(0)+E_{I_y}(0)+E_{\bar I_y}(0)  \\
	&=8(\cos\theta_x+\cos \theta_y)\(\frac{\phi}{2\pi }\)^{1/2}\re^{-A/\phi} \ .
\ea
\label{eq:E-1inst}
\ee
We will see that it indeed agrees with the numerical results given in \eqref{eq:E-1inst-num}.

\subsection{Comparison with numerical analysis}
In the previous subsection, we gave the path integral analysis in the Harper-Hofstadter model.
The spectrum of the Harper-Hofstadter Hamiltonian turned out to receive non-perturbative corrections in the weak flux limit.
The eigen-energy takes the form of the trans-series expansion \eqref{eq:E-transseries}.
As was seen in the previous subsection, the one-instanton sector consists of $x$- and $y$-instantons and their anti-instantons. From the isotropy, it should take the form as
\be
E_{(\theta_x,\theta_y)}^\text{1-inst}(N)=(\cos \theta_x+\cos \theta_y) \cN^{(1)}(N,\phi) \re^{-A/\phi}  \cP^\text{1-inst}_\text{fluc}
\label{E-1inst-general}
\ee
where $A$ is the instanton action given by \eqref{eq:A-inst}, $\cN^{(1)}(N,\phi)$ is an unknown coefficient,
and $\cP^\text{1-inst}_\text{fluc}$ is the perturbative fluctuation around the one-instanton saddle.
We have computed $\cN^{(1)}(0,\phi)$ in the previous subsection, but it is not easy to compute it for excited states in the same way.
In this subsection, we predict $\cN^{(1)}(N,\phi)$ and $\cP^\text{1-inst}_\text{fluc}$ from the numerical analysis.

To extract one-instanton contribution, the best way is probably to investigate the width of the energy bands.
As is clear from the trans-series form of the eigen-energies \eqref{eq:E-transseries} with \eqref{E-1inst-general}, 
the band width is controlled, at the leading order, by the one-instanton sector.
From the angle dependence of \eqref{E-1inst-general}, one can easily see that
two band edges correspond to $(\theta_x, \theta_y)=(0,0)$ and $(\theta_x, \theta_y)=(\pi,\pi)$.
Therefore the band width is given by
\be
\ba
	\Delta E^\text{band}(N):=|E_{(0, 0)}(N)- E_{(\pi, \pi)}(N)| 
	=4 \re^{-A/\phi} |\cN^{(1)}(N,\phi)| \cP^\text{1-inst}_\text{fluc}+\cdots
\ea
\ee
where $\cdots$ represents the higher instanton contributions, which are irrelevant here.
We compute this band width for various values of $\phi=2\pi/Q$ and $N$ using the exact formula \eqref{eq:Fab-simp}, and fix unknown parameters in the formula above by numerical fitting.
The strategy is the same as the one used in \cite{Hatsuda:2017dwx}. We refer the reader to this work for details.

We first confirm that the exponential decay of the band width in $\phi \to 0$ is actually explained by the instanton action $A=8\mathrm{C}$.
By the numerical fitting,
we also find the explicit form of the coefficient $\cN^{(1)}(N,\phi)$:
\be
\cN^{(1)}(N,\phi)=(-1)^N \frac{8^{N+1}}{\pi^N N!} \(\frac{\phi}{2\pi}\)^{\tfrac{1}{2}-N}
\label{eq:E-1inst-num}
\ee
For the lowest Landau level $N=0$, this is indeed in agreement with the path integral result \eqref{eq:E-1inst}.

Finally, we make a comment that with numerical calculation we can also go beyond the leading order contribution. With the help of the Richardson transformation, as explained in \cite{Hatsuda:2017dwx}, we find the fluctuation in the 1-instanton sector to be
\be\label{eq:P-1inst}
\ba
	\log \cP_\text{fluc}^\text{1-inst}&=
	-\frac{6N^2+30N+19}{96}\phi-\frac{20N^3+102N^2+136N+27}{4608}\phi^2 \\
	&\quad -\frac{210N^4+1380N^3+2910N^2+2700N+893}{368640}\phi^3+\cO(\phi^4) \ .
\ea
\ee

While checking this result is very hard with instanton calculus of path-integrals, our topological string theory analysis of section~\ref{sc:flucturation} essentially obtains the same result.

\section{Two-instanton sector}
\label{sc:2-inst}

\subsection{Two-instanton calculation}
\label{sc:int-anti}

Now we wish to go beyond the dilute instanton gas approximation, and compute the contributions of the two-instanton sector to the leading order in semi-classics. Recall that we have two types of instantons, which we will call $I_x$ and $I_y$, where $I_x$ is a tunneling event in $x$, i.e.\ it takes $x\rightarrow x+2\pi$, while $I_y$ is a tunneling event in $y\rightarrow y+2\pi$. 

We will consider all kinds of two-instanton events, ranging from ``pure'' correlations
\begin{equation}\label{eq:pure}
\begin{aligned}
&[I_{x}\bar I_{x}], [\bar I_{x}I_{x}], [I_{y}\bar I_{y}], [\bar I_{y} I_{y}],\\
&[I_{x}I_{x}], [I_{y}I_{y}], [\bar I_{x}\bar I_{x}], [\bar I_{y}\bar I_{y}],
\end{aligned}
\end{equation}
to ``mixed'' ones
\begin{equation}\label{eq:mixed}
\begin{aligned}
&[I_{x}I_y], [I_{y}I_x], [\bar I_x \bar I_y], [\bar I_y\bar I_x],\\
&[\bar I_x I_y], [I_y\bar I_x], [I_x\bar I_y], [\bar I_y I_x] \ .
\end{aligned}
\end{equation}
Before computing their interactions, we should stress that the contribution of such events has long been subject to debates. Particularly tricky is the instanton--anti-instanton contribution $[I \bar{I}]$, which is a priori ill-defined. This is because when instanton and anti-instanton are close to each other the configuration is indistinguishable from the perturbative vacuum, and it is not clear how such configurations should be taken into account (see \cite{Behtash:2018voa} for an incomplete list of references on the topic). 

If we naively superpose the well-separated instanton and anti-instanton, where we label their separation by $\tau$, the action will be an increasing function of $\tau$. Such a configuration spends most of the time in one of the vacua (say $x=0$) and then tunnels to the other vacuum ($x=2\pi$), lingering there for the time $\tau$, and then returns back to the original ($x=0$) vacuum. The action of such a configuration is approximately (as we will demonstrate in the next subsection)
\be\label{eq:2 instanton interaction}
	S_2 \approx 2A+ B\, \re^{-\tau}
\ee
where the exponential contribution is the ``classical'' interaction\footnote{The term ``classical'' is used to reflect the $1/\phi$ dependence of the interaction, but it is a bit of a misnomer, because an instanton--anti-instanton event is in fact a large-quantum fluctuation, and is in no way classical.} of the instanton--anti-instanton pair. The contribution of such a class of configurations to the partition function would then be\footnote{The subtracted unity is to control the $IR$ divergence due to the uncorrelated instantons. Since uncorrelated instantons have already been taken into account by the instanton gas approximation it should be subtracted here to avoid double counting.}
\be
\int \rd t_0 \,\int \rd\tau K^2\, \re^{-\frac{2A}{\phi}} \left( \re^{-\frac{B}{\phi}e^{-\tau}} -1\right)\;,
\ee
where $\phi$ is the coupling constant, $t_0$ is the ``center of mass'' location of the pair, and $K$ is the one-loop measure of the individual (anti-)instantons. The integral over $t_0$ will simply produce one power of $\beta$, while the rest of the expression will be related to the $I\bar I$ contribution to the energy. The integral over their separation is, however, an awkward operation. As we shall see in the next subsection \eqref{eq:Bxxbar}, the interaction constant $B$ is negative, so the integral is saturated by its lower limit $\tau\sim 0$, where the approximations of the above expression are invalid, and where the notion of the instanton--anti-instanton is ill defined.

Bogomolny \cite{BOGOMOLNY1980431} and Zinn-Justin \cite{ZINNJUSTIN1981125,ZinnJustin:2002ru} argued long time ago that the ill-defined $I\bar I$ amplitude is connected with the ambiguity of the Borel sum of the perturbation theory. They correctly argued that the definition of the $I\bar I$ amplitude must be ambiguous in the same way that the perturbation theory is. A prescription which is now dubbed the Bogomolny--Zinn-Justin (BZJ) prescription, is to take the coupling $\phi$ to be negative, so that the above integral is saturated away from $\tau\sim \log(1/\phi)\gg 1$, where the approximations are valid. The above integral over $\tau$ is then performed to produce a correction to the energy
\be\label{eq:generalcor}
	E_0^{I\bar I}=\phi K^2 \re^{-2A/\phi}(-\gamma_{E}- \log(B/\phi) -\Gamma(0,B/\phi))\ ,
\ee
where $\gamma_E$ is the Euler's constant, and $\Gamma(\bullet,\bullet)$ the incomplete gamma function.
The last term is exponentially small when $\phi<0$ so it is normally dropped.
Further the expression is ambiguous if we now send $\phi$ from negative to positive values in the upper or lower complex half-plane, because of the appearance of the $\log$. Moreover the ambiguity is exactly canceled by the ambiguity in the Borel sum of the perturbation theory. This was one of the great successes of resurgence in quantum mechanics and our understanding of its relationship with path-integrals.

The BZJ prescription, however revolutionary, causes some unease. Perhaps the most uncomfortable one is that it requires dropping a factor which is exponentially small when $\phi<0$, but becomes exponentially large when the correct limit $\phi>0$ is taken. In recent years it became increasingly evident that at the heart of the correct interpretation of the BZJ result is the Picard-Lefschetz theory -- a generalization of the steepest decent method to multi-integral (or indeed path-integral) cases. In fact it was only recently that a resolution of this puzzle was proposed by the interpretation of the instanton--anti-instanton pair as a saddle point at infinity \cite{Behtash:2018voa}, which establishes a concrete method for a systematic calculation of the semi-classical expansion in path integrals.
The procedure is roughly as follows (We refer the reader to \cite{Behtash:2018voa} for details.):
\renewcommand\labelenumi{\arabic{enumi})}
\begin{enumerate}
	\item Consider an instanton--anti-instanton configuration for the case of finite time $\beta$.
	\item Note that if the instanton and the anti-instanton are at opposite ends of a temporal circle, the configuration becomes a saddle point. Since the action can be decreased by bringing the pair closer together, the saddle point in question is ``unstable''.
	\item Treat the saddle point with Picard-Lefschetz theory, i.e.\ instead of integrating over a cycle of real instanton--anti-instanton separation, replace the cycle with the Lefschetz thimble integral (i.e.\ the ``steepest decent cycle"), along which the action is monotonically increasing.
	\item Note that the imaginary part of the thimble integral is ambiguous depending on whether $\text{Im }\phi$ is greater or smaller than zero, and that the ambiguity cancels the Borel sum ambiguity of the path-integral, while the real part is identical to the BZJ result above, provided that we drop the incomplete-gamma term. 
\end{enumerate}

In particular the ambiguity, which comes from the imaginary part, is given by
\be\label{eq:ImE}
	\text{Im }E_0^{I_x \bar I_x}=\pm \pi\phi\, K^2 \, e^{-2 A/\phi}=\pm  8\, \re^{-2A/\phi}
\ee
where we used our result \eqref{eq:K}. 
We would like to point out that the ambiguity does not contain the interaction term for the ground-state energy, i.e.\ it is independent of the constant $B$ which parametrizes the instanton--anti-instanton interactions. This is in fact clarified by the thimble integration procedure in \cite{Behtash:2018voa}, summarized above. The ambiguity comes from the vicinity of the critical point at infinity, which, for a finite temporal extent, is the instanton--anti-instanton pair at opposing ends of the temporal circle. Since the saddle is ``unstable'' with regards to the perturbations in the real field space, the proper thimble integration will force us to integrate along the direction of imaginary separation\footnote{The contour $I\bar I$ separation parameter $\tau$ along the thimble eventually bends and becomes parallel to the real axis in the complex $\tau$-plane, which gives the real contribution.}, inducing an imaginary factor in the result. This is the ambiguity, and in this case it is saturated in the vicinity of the $I\bar I$ saddle. When we take $\beta\rightarrow \infty$, this vicinity of the $I\bar I$ saddle moves to infinity, where the instanton and anti-instanton are decorrelated, and all dependence on the interactions vanishes. 

On the other hand, the real part is given by
\be\label{eq:EIbarI}
	\text{Re }E_0^{I_x\bar I_x}=\phi\, K^2\, \re^{-2A/\phi}(-\gamma_E-\log(-B/\phi)) = \frac{8}{\pi}\, \re^{-2A/\phi}(-\gamma_E-\log(-B/\phi))\;,
\ee
which depends explicitly on the instanton interaction parameter $B$, and which compared to corresponding terms in \eqref{eq:generalcor} has already changed sign inside the logarithm. We leave the computation of the interaction parameter to the next subsection.

Let us now consider other two-instanton events with nonzero topological charges. As opposed to quantum mechanics these in addition to the pure types (second line in \eqref{eq:pure}) include also the mixed types \eqref{eq:mixed}. It is straightforward to repeat the same analysis as for the instanton--instanton events, and it yields always more or less the same results.
However, a crucial difference is that, as we shall see in the next subsection, the interaction constant $B$ is positive if the two instantons are identical. In this case we have no need to change the sign of the coupling in the logarithm and the ambiguity is absent. The energy correction contains the real part only, which is \eqref{eq:EIbarI} with $B$ replaced by $-B$. Furthermore the mixed type as in \eqref{eq:mixed} also must be present on physical grounds, because we expect terms of the kind $\cos(\theta_x)\cos(\theta_y)$ to be present in the ground-state energy $\theta$-dependence. Quite unexpectedly their interactions are all found to be purely imaginary, and thus the individual corresponding energy correction is complex. Nevertheless, when all eight mixed 2-instanton events in \eqref{eq:mixed} are considered, the total energy correction is real.

In fact, according to the logic of \cite{Behtash:2018voa} all these contributions should correspond to exact saddles of the QM problem on a compact $S^1$ time. What is especially interesting is that while the instanton-instanton and instanton--anti-instanton events of the same type \eqref{eq:pure} have their counterpart in quantum mechanics and are thus not very surprising by analogy\footnote{Such saddles can be thought of as the motions in a periodic inverted-potential which either oscillate with a period $\beta$ between two peeks of the inverted potential (i.e.\ between two classical vacua of the potential) -- a saddle  that corresponds to an instanton--anti-instanton pair -- or roll with the ``energy'' slightly higher than the peak of the inverted potential so that in precisely one period of the imaginary time $\beta$ the particle winds twice -- a saddle corresponding to an instanton-instanton event.}, the mixed type \eqref{eq:mixed} are a different matter. Yet as we shall see their naive BZJ amplitudes are in an extremely good agreement with the numerics, so we are inclined to believe that such saddles must also exist.

\subsection{Instanton interactions}

This section is devoted to the determination of instanton-interaction constants $B$ in various instanton events, and the corresponding correction to the ground state energy. To start with, we would like to write down a general formula with any given choice of correlations. The ansatz goes as follows: Let's consider a superposition of two instanton events,
\be
x_{2} = \xa + \xb\, ,\ y_{2} = \ya + \yb\, ,
\ee
where $\xa$ and $\xb$ ($\ya$ and $\yb$) are either instanton or anti-instanton in the $x\  (y)$ direction. We shift the solutions to separate two events $(\xa, \ya)$ and $(\xb, \yb)$ by $\tau \gg 0$. We further make the assumption that the ``tunneling'' of $(\xa, \ya)$ takes place when $t \ll 0$, while for $(\xb, \yb)$ it happens when $t \gg 0$. As a consequence, it shows that $\xa(0)$ or $\ya(0)$ differs from $\xa(+\infty)$ or $\ya(+\infty)$ by exponentially small terms, while $\xb(0)$ or $\yb(0)$ differs from $\xb(-\infty)$ or $\xb(-\infty)$ by exponentially small terms.

The two-instanton action can be split into two parts,
\be\label{eq:lagrangian}
	S_2 = \int_{-\infty}^{\infty} \rd t\ \mathcal{L}(x_{2},y_{2}) = \int_{-\infty}^{0} \rd t\ (H(x_{2},y_{2}) - \ri \dot{x}_{2} y_{2})
+ \int_{0}^{\infty} \rd t\ (H(x_{2},y_{2}) - \ri \dot{x}_{2} y_{2})\ .
\ee
For ease of notation, we denote the two terms on the r.h.s.\ as $S_{-}$ and $S_{+}$ respectively. 

Let's first concentrate on $S_{-}$. Our assumption implies that $\dex = \xb - \xb|_{-\infty}$ and $\dey = \yb - \yb|_{-\infty}$ can be treated as small perturbations in this region. Let us Taylor expand $S_-$ up to the second order
\be
\ba
S_{-} = &\int_{-\infty}^{0} \rd t\ ( H|_{\xa,\ya} + \partial_{x}H|_{\xa}\dex + \partial_{y}H|_{\ya}\dey + \frac{1}{2}\partial^{2}_{x}H|_{\xa}\dex^2 + \frac{1}{2}\partial^{2}_{y}H|_{\ya}\dey^2 \\
&- \ri\dot{x}_{\alpha} (\ya + \yb|_{-\infty}) - \ri \dot{x}_{\alpha} \dey - \ri \delta \dot{x} (\ya + \yb|_{-\infty}) - \ri\delta\dot{x}\dey)\ .
\ea
\ee
Here we have used the fact that both $\xb(-\infty)$ and $\yb(-\infty)$ must be integer multiples of $2\pi$. 
It can be shown that sum of the two terms $\int_{-\infty}^{0} \rd t (\frac{1}{2}\partial^{2}_{x}H|_{\xa}\dex^2 + \frac{1}{2}\partial^{2}_{y}H|_{\ya}\dey^2)$ always has the order $o(e^{-2\tau})$, while we will look for the interactions of order $o(e^{-\tau})$. Indeed the reader may wonder why we even bothered to expand up to a quadratic order in $\delta x$ and $\delta y$ in the first place. The reason is that the terms $\delta y\delta \dot x$ is special because it is a total derivative, and will contribute a finite amount to the $o(e^{-\tau})$ order. Then recall our normalization and  the equations of motion \eqref{eq:eom1}\eqref{eq:eom2} as well as the conservation law \eqref{eq:inst1-Esimp}, we can further simplify it to be 
\be
- \ri  \dex(0) \left(\ya|_{0} + \yb|_{-\infty}\right) - \ri\int_{-\infty}^{0}\rd t\ \delta\dot{x} \dey  - \ri \yb|_{-\infty} \int_{-\infty}^{0}\rd t\ \dot{x}_{\alpha}  -\ri\int_{-\infty}^{0}\rd t\ \dot{x}_{\alpha}y_{\alpha}\ .
\ee
If we choose the normalization carefully such that $\yb|_{-\infty} = 0$ and $- \ri \int_{-\infty}^{\infty}\rd t\  \dot{x}_{\alpha}y_{\alpha} = A$, we can obtain, at the leading order, a very neat formula
\be
S_{-} = A + \ri (\xb|_{-\infty} - \xb|_{0}) (\ya|_{0} + \yb|_{0}) + \ri \ya|_{0}\left(\xa|_{\infty} - \xa|_{0}\right)\, .
\ee

The same story, with the only difference that $\dpx = \xa - \xa|_{+\infty}$ and $\dpy = \ya - \ya|_{+\infty}$ are treated as small perturbations, goes for $S_{+}$ and yields
\be
S_{+} = A - \ri (\xa|_{\infty} - \xa|_{0}) (\ya|_{0} + \yb|_{0}) - \ri \yb|_{0}\left(\xb|_{-\infty} - \xb|_{0}\right)\, ,
\ee
given the choice of renormalization $\ya|_{\infty} = 0$ and $- \ri \int_{-\infty}^{\infty}\rd t\ \dot{x}_{\beta}y_{\beta} = A$ (again we observe that $\int^{\infty}_{0} \rd t (\frac{1}{2}\partial^{2}_{x}H|_{\xb}(\dpx)^2 + \frac{1}{2}\partial^{2}_{y}H|_{\yb}(\dpy)^2)$ has no contribution at this order).

Summing up $S_+$ and $S_-$ we get the following approximation of two-instanton action
\be\label{eq: interaction}
	S_2 = 2A + \ri\ya|_{0}\left(\xb|_{-\infty} - \xb|_{0}\right) - \ri \yb|_{0}\left(\xa|_{\infty} - \xa|_{0}\right)\ .
\ee
Notice that $2 A$ is already accounted for by dilute instanton gas approximation, while the remaining part will yield the exponential contribution predicted in \eqref{eq:2 instanton interaction}.

Now it is time to consider concrete examples and plug in instanton solutions. First of all, it is convenient to recall the asymptotic behaviors of instanton solutions
\be\label{eq:asymtotics}
x_{1}(t) = \begin{cases} 2\sqrt{2} e^{-t}, \ t \ll 0 \\ 2\pi - 2\sqrt{2} e^{-t},\ t \gg 0\end{cases}, \ \ y_{1}(t) = \begin{cases} \ri 2\sqrt{2} e^{-t}, \ t \ll 0 \\ \ri 2\sqrt{2} e^{-t},\ t \gg 0\end{cases}.
\ee
Let us go over all the 2-instanton events listed in \eqref{eq:pure},\eqref{eq:mixed}.

\begin{itemize}
	\item ``Pure'' instanton--anti-instanton: The four events $[I_{x}\bar I_{x}], [\bar I_{x}I_{x}], [I_{y}\bar I_{y}], [\bar I_{y}I_{y}]$ in the first line of \eqref{eq:pure} have the same interaction term, so we only need to compute $[I_{x}\bar I_{x}]$. A superposition can be chosen as
	\be
	\ba
		\xa &= x_{1}(t + \frac{\tau}{2})\, ,\  \xb = x_{1}(-t +\frac{\tau}{2}) - 2\pi\, , \\ 
		\ya &= y_{1}(t + \frac{\tau}{2})\, ,\  \yb = -y_{1}(-t + \frac{\tau}{2})\, .\\
	\ea
	\ee
	The first order interaction can be read off with the help of asymptotics \eqref{eq:asymtotics} and we obtain
	\be\label{eq:S2-pure1}
		S_{ I_x\bar{I}_x } = 2A - 16\, \exp(-\tau)\, ,
	\ee
	which verifies the claim \eqref{eq:2 instanton interaction} and we find the instanton-interaction constant
	\begin{equation}\label{eq:Bxxbar}
		B_{I_x\bar{I}_x} = -16
	\end{equation}
	to be negative.

	\item ``Pure'' instanton-instanton: The four events in the second line of \eqref{eq:pure} also have the same interaction, so we only need to compute $[I_{x}I_{x}]$. A superposition can be chosen as 
	\be
	\ba
		\xa &= x_{1}(t + \frac{\tau}{2})\, ,\  \xb = x_{1}(t -\frac{\tau}{2})\, , \\ 
		\ya &= y_{1}(t + \frac{\tau}{2})\, ,\  \yb = y_{1}(t - \frac{\tau}{2})\, .\\
	\ea
	\ee
	We readily find
	\be\label{eq:S2-pure2}
		S_{ I_x I_x} = 2A + 16\, \exp(-\tau)\, ,
	\ee
	from which we read off the instanton-interaction constant
	\begin{equation}\label{eq:Bxx}
		B_{I_xI_x} = 16 \ ,
	\end{equation}
	which is positive.

	\item ``Mixed'' events: We work out explicitly the $[I_{x}I_{y}]$ pair. A superposition satisfying the constraints above can be chosen as
	\be
	\ba
		\xa &= x_{1}(t + \frac{\tau}{2})\, ,\  \xb = y_{1}(t -\frac{\tau}{2})\, , \\ 
		\ya &= y_{1}(t + \frac{\tau}{2})\, ,\  \yb = - x_{1}(t - \frac{\tau}{2})\, .\\
	\ea
	\ee
	Plugging into our general formula \eqref{eq: interaction}, we get
	\be\label{eq:S2-mix1}
		S_{ I_{x} I_{y}} = 2A + 16\, \ri\, \exp(-\tau)\, .
	\ee
	On the other hand, if we consider the $[I_{y}I_{x}]$ correlation, we need to shift our solutions
	\be
	\ba
		\xa &= y_{1}(t + \frac{\tau}{2})\, ,\  \xb = x_{1}(t -\frac{\tau}{2})\, , \\ 
		\ya = &2\pi - x_{1}(t + \frac{\tau}{2})\, ,\  \yb = y_{1}(t - \frac{\tau}{2})\, ,\\
	\ea
	\ee
	thus we obtain
	\be\label{eq:S2-mix2}
		S_{ I_{y} I_{x}} = 2A - 16\, \ri\, \exp(-\tau)\, .
	\ee
	Indeed these 2-instanton actions share the same pattern as \eqref{eq:2 instanton interaction} but with imaginary instanton-interaction constants. By the same token, we are able to determine all the rest ``mixed'' events
	\be\label{eq:mixedresult}
	\ba
		S_{ I_{x} \bar{I}_{y}} &= 2A - 16\, \ri\, \exp(-\tau)\, ,\  S_{\bar{I}_{y}I_{x}} = 2A + 16\, \ri\, \exp(-\tau)\\
		S_{ \bar{I}_{x} I_{y}} &= 2A - 16\, \ri\, \exp(-\tau)\, ,\  S_{I_{y}\bar{I}_{x}} = 2A + 16\, \ri\, \exp(-\tau)\\
		S_{ \bar{I}_{x} \bar{I}_{y}} &= 2A + 16\, \ri\, \exp(-\tau)\,,\  S_{ \bar{I}_{y}\bar{I}_{x}} = 2A - 16\, \ri\, \exp(-\tau) .\\
	\ea
	\ee
\end{itemize}

With all the formulas in hand, we are able to determine various contributions to the ground state energy due to various 2-instanton events. We assume that up to 2-instantons, the ground state energy of the Harper-Hofstadter model has the most general trans-series form
\begin{equation}\label{eq:ansatz}
	E_0(\theta_x,\theta_y) = E_0^{\text{pert}} + E_0^{\text{1-inst}}(\cos \theta_x + \cos \theta_y) + E_0^{I\bar{I}} + E_0^{II}(\cos 2\theta_x + \cos 2\theta_y) + E_0^{II\text{mix}} \cos \theta_x\cos \theta_y  \ ,
\end{equation}
which respects the symmetry $\theta_x \to \theta_y$ as well as $\theta_x \to -\theta_x, \theta_y \to -\theta_y$.

The 1-instanton correction has already been discussed in section~\ref{sc:1-inst}. We will check the various 2-instanton corrections in this section. Let's first look at the $E_0^{I\bar{I}}$ term. From the discussion above, we know that this term has both real and imaginary parts, given by \eqref{eq:EIbarI} and \eqref{eq:ImE} respectively for an individual 2-instanton event. Reading off the instanton interaction constant $B_{I_x\bar{I}_x}$ from \eqref{eq:S2-pure1}, and summing up all four events in the first line of \eqref{eq:pure}, we find the real correction is
\be\label{eq:E-2inst-pure1}
	\text{Re } E_0^{I\bar{I}} = \re^{-2A/\phi}\frac{32}{\pi}\(\gamma_E + \log(16/\phi)\).
\ee
while the imaginary correction, i.e.\ the ambiguity is
\be\label{eq:ambE}
	\text{Im } E_0^{I\bar{I}} =\pm 32\, \re^{-2A/\phi}\ ,
\ee

Next, $E_0^{II}$ receives contribution from both $I_{x}I_{x}$ and $\bar{I}_x \bar{I}_x$ events, which is the same as the sum of $I_yI_y$ and $\bar{I}_y\bar{I}_y$. Since the instanton-interaction $B_{I_xI_x}$ \eqref{eq:Bxx} is negative, the energy correction is real, and we find
\be\label{eq:E-2inst-pure2}
	E_0^{II} = \re^{-2A/\phi}\frac{16}{\pi}\(\gamma_E + \log(16/\phi)\).
\ee

Finally, all the eight mixed events listed in \eqref{eq:mixed} contribute to $E_0^{II\text{mix}}$. Although each individual event has imaginary instanton-interaction, as one sees in \eqref{eq:S2-mix1},\eqref{eq:S2-mix2},\eqref{eq:mixedresult}, and thus gives complex correction to the ground state energy, one can check that the imaginary contributions cancel and the total contribution of all the eight events is real. It reads
\be\label{eq:E-2inst-mix}
	E_0^{II\text{mix}} = \re^{-2A/\phi}\frac{64}{\pi}\(\gamma_E + \log(16/\phi)\).
\ee

\subsection{Numerical studies of two-instanton sector}
\label{sc:n2inst}

\begin{figure}[t]
	\centering
	\subfloat[$E_0^{II}$]{\includegraphics[width=0.4\linewidth]{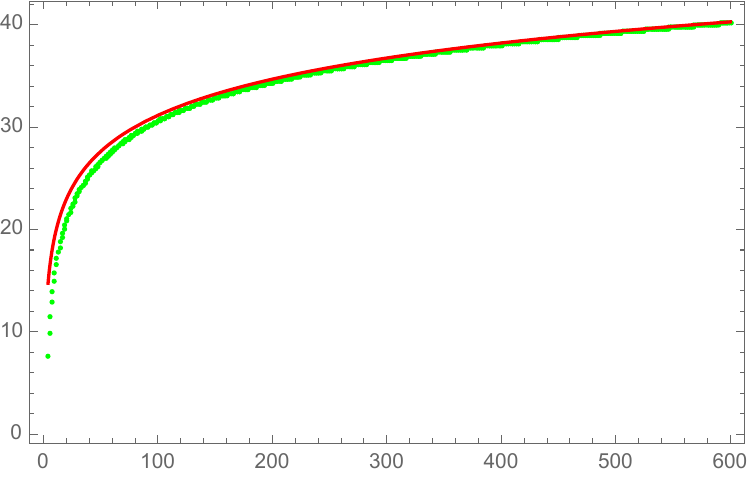}}\hspace{2ex}
	\subfloat[$E_0^{I\bar{I}}$]{\includegraphics[width=0.4\linewidth]{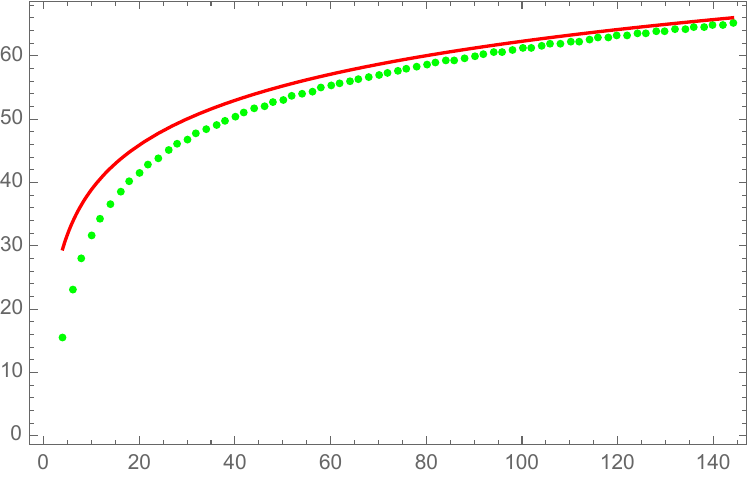}}\\
	\subfloat[$E_0^{II}$]{\includegraphics[width=0.4\linewidth]{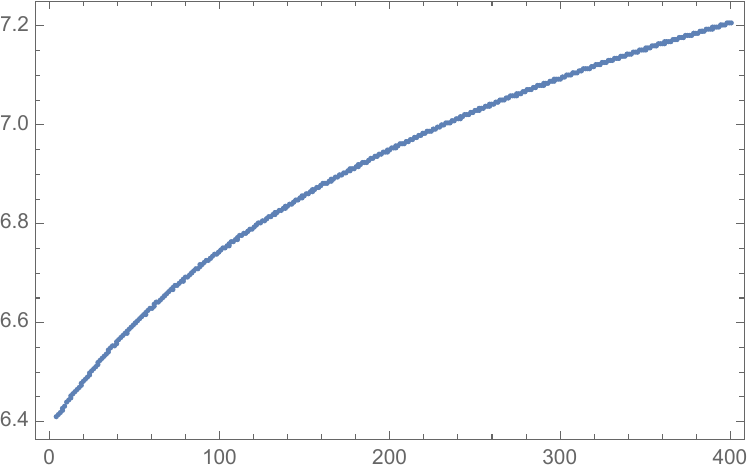}}\hspace{2ex}
	\subfloat[$E_0^{I\bar{I}}$]{\includegraphics[width=0.4\linewidth]{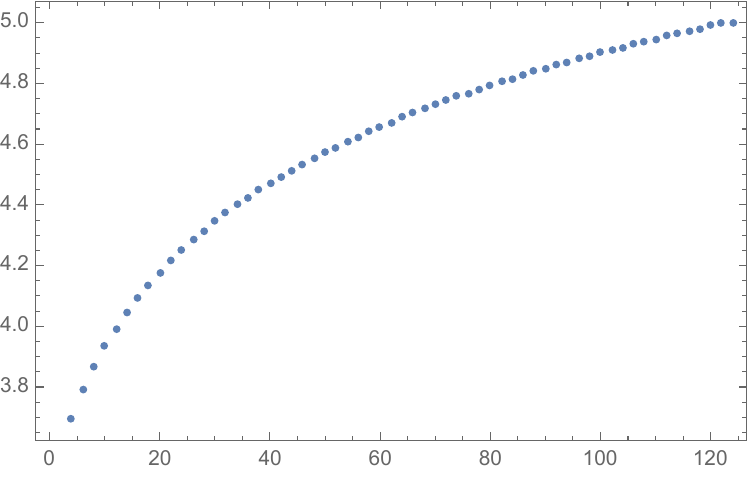}}
	\caption{We plot in upper panels numerical results of 2-instanton corrections (left: $E_0^{II}$; right: $E_0^{I\bar{I}}$) as a function of $Q = 2\pi/\phi$ in green dots versus theoretical predictions \eqref{eq:E-2inst-pure1}, \eqref{eq:E-2inst-pure2} in red lines. Plotted in lower panels are the matching digits between the numerical and theoretical results; we perform Richardson transformations of order 200 for $E_0^{II}$ and order 10 and $E_0^{I\bar{I}}$ respectively to improve convergence.}\label{fg:Ixx-Ixy}
\end{figure}

In this subsection, we perform a numerical study of the various 2-instanton corrections to the ground state energy, and compare them with the predictions computed in the last subsection. And we will find perfect agreement. We confine ourselves to the real parts of the corrections, and leave the study of the imaginary part (ambiguity) to the next subsection. 

The trans-series \eqref{eq:ansatz} of the ground state energy already gives us a hint as how to extract 2-instanton corrections numerically. We have
\begin{equation}
\begin{aligned}
	&E_0(0,\tfrac{\pi}{2}) = E_0^\text{pert} + E_0^{\text{1-inst}} + E_0^{I\bar{I}} \ , \\
	&E_0(0,\pi) = E_0^\text{pert} + 2E_0^{II} - E_0^{II\text{mix}} + E_0^{I\bar{I}} \ ,\\
	&E_0(\tfrac{\pi}{2},\tfrac{\pi}{2}) = E_0^\text{pert} - 2 E_0^{II} + E_0^{I\bar{I}} \ ,\\
	&E_0(\tfrac{\pi}{2},\pi) = E_0^\text{pert} - E_0^{\text{1-inst}} + E_0^{I\bar{I}} \ ,
\end{aligned}
\end{equation}
thus all the 2-instanton contributions can be obtained from the following linear contributions
\begin{equation}\label{eq:2inst-lin}
\begin{aligned}
	E_0^\text{pert} + E_0^{I\bar{I}} &= \frac{1}{2}( E_0(0,\tfrac{\pi}{2}) + E_0(\tfrac{\pi}{2},\pi)) \ ,\\
	E_0^{II} &= \frac{1}{4} (E_0(0,\tfrac{\pi}{2}) + E_0(\tfrac{\pi}{2},\pi) - 2 E_0(\tfrac{\pi}{2},\tfrac{\pi}{2})) \ ,\\
	E_0^{II\text{mix}} &= E_0(0,\tfrac{\pi}{2}) + E_0(\tfrac{\pi}{2},\pi) -E_0(\tfrac{\pi}{2},\tfrac{\pi}{2}) - E_0(0,\pi) \ .
\end{aligned}
\end{equation}
Since the r.h.s.\ can be computed exactly when $\phi = 2\pi/Q$ for $Q\in \mathbb N$, these simple linear formulas allow us to easily compute 2-instanton corrections indicated on the l.h.s.\ up to at least 3-instanton corrections for a sequence of $Q$ up to very large $Q$, with very good accuracy for large $Q$. Note that here $E_0^{I\bar{I}}$ actually refers to only its real part; the imaginary value cannot be computed in this way as it cancels in physical observables. To check the imaginary part of the 2-instanton sector we can analyze the perturbation series, and match its lateral Borel sum with the ambiguity from the instantons. This will be discussed in the next section. We also notice that since $E_0(\pi/2,\pi/2) = E_0(0,\pi)$ (c.f.\ \eqref{eq:Fab-simp}), we have
\begin{equation}\label{eq:IImix-II}
	E_0^{II\text{mix}} = 4E_0^{II} \ ,
\end{equation}
up to the next instanton level, which is indeed implied by \eqref{eq:E-2inst-pure2},\eqref{eq:E-2inst-mix}. Thus we can skip $E_0^{II\text{mix}}$ and only check $E_0^{II}$ if \eqref{eq:2inst-lin} are correct.

We comment that in using \eqref{eq:2inst-lin} we will make at most an error exponentially suppressed by a one-instanton factor. We demonstrate this explicitly by comparing with the results of Fourier transformation in Appendix~\ref{app:fourier}

Let us thus 
focus on $E_0^{I\bar{I}}$ and $E_0^{II}$. For a sequence of $Q=2\pi/\phi$, we expect improving agreement with the path integral predictions \eqref{eq:E-2inst-pure1},\eqref{eq:E-2inst-pure2} as $Q$ increases. Note that from numerics, we only obtain the combination $E_0^\text{pert}+E_0^{I\bar{I}}$, and we have to remove $E_0^\text{pert}$ by hand by subtracting the Borel-Pad\'e sum of the perturbative ground state energy. Poles in the Borel plane, which are responsible for the ambiguity \eqref{eq:ambE}, are dealt with by Cauchy principal value integration. This additional complication limits the range of $Q$ we can push for. The agreement between the numerical results and the path integral predictions is excellent, as demonstrated in the matching digits plots Fig.~\ref{fg:Ixx-Ixy}.

\subsection{Large order growth and ambiguity of energy}
\label{sc:growth}

According to the resurgence theory, the large order growth of the perturbative energy expansion is controlled by the ambiguity (imaginary part) of energy, which receives contributions from instanton sectors with topological charge zero (see for instance \cite{Dorigoni:2014hea}). The first such sector is the instanton--anti-instanton sector $[I\bar{I}]$ including all four events listed in the first line of \eqref{eq:pure}. The imaginary energy correction from this sector is
\begin{equation}
	\text{Im}E^{I\bar{I}}(N,\phi) = \pm  \,\re^{-2A/\phi} (S_{(N)}/2) \cdot \phi^{b_N}\sum_{n=0}^{\infty} a_n^{(1,1)}(N) \phi^n \nn
\end{equation}
where $S_{(N)}$ is the Stoke's constant, related to the ambiguity of the lateral Borel resummation of the perturbative expansion, and $b_N$ is the leading exponent of $\phi$ in the instanton--anti-instanton sector. Let us denote the perturbative expansion by
\be
E^\text{pert}(N)=\sum_{n=1}^\infty a_n^{(0)}(N) \phi^n
\ee
The resurgent analysis then suggests the following relation
\be
	a_n^{(0)}(N)=\frac{S_{(N)}}{2\pi} \frac{(n-b_N-1)!}{ (2A)^{n-b_N}}
	\( 1+\frac{a_1^{(1,1)}(N) 2A}{n-b_N-1}+\frac{a_2^{(1,1)}(N) (2A)^2}{(n-b_N-1)(n-b_N-2)}+\cdots \)\ .
	\label{eq:aNn}
\ee
We will use this relation to compute numerically the imaginary part of $E^{I\bar{I}}$.

We start with the ground state with $N=0$.
We compute $a_n^{(0)}$ up to $n=320$ using the \texttt{BenderWu} package. With the help of \eqref{eq:aNn}, we found that
\begin{equation}
	b_0 = 0 \ ,
\end{equation}
and we also extracted the following numerical values of $A$ and $S_{(0)}$
\be
\ba
	2A^\text{num}     =14.6554495068355\dots \ ,\qquad
	S_{(0)}^\text{num}=63.9999999999999\dots \ .
\ea
\ee
In this process, it is convenient to use the Richardson transformation to accelerate the convergence (see for instance \cite{Marino:2007te} for details). It is easy to check that these numerical estimations reproduce the exact values
\begin{equation}\label{eq:S}
	2A=16\Catalan,\qquad S_{(0)} = 64\ ,
\end{equation}
so that in the leading order, we have
\begin{equation}
	\text{Im} E^{I\bar{I}} (0,\phi) = \pm  32\re^{-16\Catalan/\phi} \ ,
\end{equation}
which agrees with the path integral calculation \eqref{eq:ambE}.

As in the 1-instanton sector, once the analytic values of $S_{(0)}$ and $A$ are fixed, numerically we can go beyond the leading order and further extract the values of  $a_{n}^{(1,1)}(0)$ using \eqref{eq:aNn}. For instance, we find
\begin{equation}
\begin{gathered}
	a_{1}^{(1,1)}(0)=-\frac{13}{48}\ ,\quad
	a_{2}^{(1,1)}(0)=\frac{115}{4608}\ ,\\
	a_{3}^{(1,1)}(0)=-\frac{12209}{3317760}\ ,\quad
	a_{4}^{(1,1)}(0)=-\frac{355687}{637009920}\ ,\cdots
\end{gathered}
\end{equation}
These coefficients should give the perturbative fluctuation around the instanton--anti-instanton saddle.

We repeat the same computation for higher energy levels. Observing the general structure \eqref{eq:aNn},
we find that
\begin{equation}
	b_N = -2N \ ,\quad S_{(N)} = \frac{2^{8N+6}}{(N!)^2} \ .
\end{equation}
In addition, the fluctuation around the $[I\bar{I}]$ saddle point should be
\be\label{eq:P-bion}
\ba
	\log \cP_\text{fluc}^{I\bar{I}}:=&\log \( \sum_{n=0}^\infty a_n^{(1,1)}(N) \phi^n \) \\
	=&-\frac{6N^2+18N+13}{48}\phi-\frac{20N^3+66N^2+100N+27}{2304} \phi^2 \\
	&-\frac{210N^4+900N^3+2190N^2+1980N+653}{184320} \phi^3+\cO(\phi^4).
\ea
\ee
From these data, we could construct the $[I\bar{I}]$ contribution to the imaginary part of the eigen-energy 
\begin{align}\label{eq:E-bion}
	\text{Im}E^{I\bar{I}}(N,\phi) =& \pm \ri \,\re^{-2A/\phi} (S_{(N)}/2) \cdot \phi^{b_N}\sum_{n=0}^{\infty} a_n^{(1,1)}(N) \phi^n \nn
	=&\pm \ri \,\re^{-2A/\phi} \frac{2^{8N+5}}{(N!)^2} \phi^{-2N} \cdot \cP_\text{fluc}^{I\bar{I}} \ .
\end{align}

Before we conclude this section, we point out that there is an interesting relation between $\cP_\text{fluc}^{I\bar{I}}$ and $\cP_\text{fluc}^\text{1-inst}$
\begin{equation}\label{eq:threesome}
\frac{\cP_\text{fluc}^{I\bar{I}}}{(\cP_\text{fluc}^\text{1-inst})^2}
=\( \frac{1}{\phi} \frac{\pd E^\text{pert}}{\pd N} \)^{-1} \ .
\end{equation}
which indicates that we can can cast the 1-instanton fluctuation and $[I\bar{I}]$ fluctuation as
\begin{align}
	\cP_\text{fluc}^\text{1-inst} &=\frac{1}{\phi} \frac{\pd E^\text{pert}(N)}{\pd N} e^{-\cA(N,\phi)} \ , \label{eq:inst-fluc}\\
	\cP_\text{fluc}^{I\bar{I}} &=\frac{1}{\phi} \frac{\pd E^\text{pert}(N)}{\pd N} e^{-2\cA(N,\phi)}\ . \label{eq:bion-fluc}
\end{align}
where the function $\cA(N,\phi)$ is nothing else but the ``non-perturbative'' A-function appearing in the Zinn-Justin--Jentschura exact quantization conditions \cite{ZJJ1, ZJJ2} in conventional quantum mechanics. In our example, the first few terms of $\cA(N,\phi)$ read
\be
\ba
	\cA(N,\phi)=\( \frac{\nu^2}{16}+\frac{11}{192}\)\phi
	+\(\frac{5\nu^3}{1152}+\frac{49}{4608} \)\phi^2 \\
	+\( \frac{7\nu^4}{12288}+\frac{77\nu^2}{24576}+\frac{889}{2949120}\) \phi^3+\cO(\phi^4)\ .
\ea
\label{eq:A-func-num}
\ee
where $\nu=N+1/2$.

\section{Instanton fluctuation from topological string}
\label{sc:flucturation}

Here we reveal an interesting connection between the fluctuation parts  $\cP_\text{fluc}^\text{1-inst} ,\cP_\text{fluc}^{I\bar{I}}$ and topological string theory.

Before our analysis, we would like to remind the reader that the Harper-Hofstadter model is closely related to a Calabi-Yau threefold called the canonical bundle of $\bF_0$, also known as local $\bF_0$ in string theory community, as first pointed out in \cite{Hatsuda:2016mdw}. According to mirror symmetry, all the Gromov-Witten invariants of local $\bF_0$ are encoded in an algebraic curve, called mirror curve, whose equation reads\footnote{We have set one coefficient of the curve equation, the so-called mass parameter, to be 1. This mass parameter corresponds to anisotropy of the 2d lattice.}
\begin{equation}\label{eq:F0-curve}
	\re^x + \re^{-x} + \re^{y} + \re^{-y} = u \ .
\end{equation}
Clearly the Hamiltonian of the Harper-Hofstadter model \eqref{eq:Hof_Hamiltonian} can be obtained by rotating $(x,y)$ in complex plane to $(\ri x, \ri y)$, and promoting them to operators satisfying the commutation relation \eqref{eq:xyphi}. Then the free parameter $u$ is related to the energy by $u=4-2E$. One can obtain another QM model by promoting $x$ and $y$ without the rotation, i.e.,~one considers the Hamiltonian
\begin{equation}\label{eq:H-F0}
	\cH^{\bF_0} = -\frac{1}{2}\(\re^\sx + \re^{-\sx} + \re^{\sy} + \re^{-\sy}\) + 2 \ ,
\end{equation}
with 
\begin{equation}
	[\sx,\sy] = \ri \hbar \ ,\quad \hbar \in \bR_+ \ .
\end{equation}
We choose a normalization of $\cH^{\bF_0}$ slightly different from that in the literature to match the normalization of \eqref{eq:Hof_Hamiltonian} we use in this paper. 
Motivated by topological string considerations \cite{Aganagic:2003qj,Aganagic2012}, this QM model has been thoroughly studied, both its spectrum \cite{Kallen2013,Grassi:2014zfa,Kashaev:2015kha,Wang2015,Sun2016,Grassi:2016nnt,Grassi:2017qee} and its wave functions \cite{Marino:2017gyg,Marino:2016rsq,Zakany:2017txl} (see also \cite{Kashani-Poor:2016edc,Sciarappa:2016ctj}).  
This has led to exciting development of non-perturbative completion of topological string theory and topological string / spectral theory duality \cite{Grassi:2014zfa,Marino:2015ixa,Kashaev:2015wia,Marino:2015nla,Codesido2015a,Bonelli:2016idi,Bonelli:2017ptp,Bonelli:2017gdk}, which in turn inspired a new procedure to solve non-perturbatively QM models \cite{Codesido:2016dld,Codesido:2017jwp,Fischbach:2018yiu}, as well as the discovery of a new class of exactly solvable deformed QM models \cite{Grassi:2018bci}.

We would like to point out that on the one hand, 
the Hamiltonian \eqref{eq:H-F0} and that of the Harper-Hofstadter model  are rather different in nature. The former is confining and has a discrete spectrum, while the Harper-Hofstadter Hamiltonian is periodic and thus has a rich band structure. On the other hand, the spectra of the two Hamiltonians are closely related in the semi-classical regime. In fact, the perturbative eigen-energies of $\cH^{\bF_0}$ was computed in \cite{Codesido:2017jwp}, also using the \texttt{BenderWu} package \cite{Sulejmanpasic:2016fwr, Gu:2017ppx}, and it is easy to check that they are related to the perturbative eigen-energies of $\cH(0,0)$ by the map
\begin{equation}
	\hbar \to -\phi \ . \label{eq:replace}
\end{equation}
We will see in later sections that many results \cite{Codesido:2017jwp} also apply for the Harper-Hofstadter model as well with appropriate modification.

The large order growth of the perturbative energy of $\cH^{\bF_0}$ has been analyzed in detail in \cite{Codesido:2017jwp}, and it is incorporated in the leading non-perturbative correction\footnote{This is what is called the 1-instanton correction in \cite{Codesido:2017jwp}. We refer to it as the ``instanton--anti-instanton'' correction because of the similarity to the Harper-Hofstadter model. More precisely, the situation in \cite{Codesido:2017jwp} corresponds to the special Bloch angles $(\theta_x, \theta_y)=(\pi/2, \pi/2)$, which is just the midpoint (or the Van Hove singularity) of each subband. At this point, the one-instanton correction vanishes, and the leading non-perturbative correction starts from the two-instanton order.} to the perturbative series. It is revealed in \cite{Codesido:2017jwp} that this non-perturbative correction can be obtained from the refined free energies in the Nekrasov-Shatashvili limit of topological string theory on the Calabi-Yau threefold local $\bF_0$. We will demonstrate that we can obtain the 1-instanton correction (and the instanton--anti-instanton correction) of the Harper-Hofstadter model from their data by applying the map $\hbar \to -\phi$.
This is not obvious at first glance because the 1-instanton correction here is the half order of the non-perturbative correction in \cite{Codesido:2017jwp}. This is a consequence of the fact that the 1-instanton sector and the instanton--anti-instanton sector are closely interrelated, as suggested in \cite{Hatsuda:2017dwx}.

Let us  quickly review the results of \cite{Codesido:2017jwp} concerning the spectrum of $\cH^{\bF_0}$. The perturbative eigen-energy can be computed also by using the \texttt{BenderWu} package \cite{Sulejmanpasic:2016fwr,Gu:2017ppx}, and the first few terms read
\begin{equation}
	E^\text{pert}_{\bF_0}(\nu,\hbar) = -\nu\hbar - \frac{4\nu^2+1}{32}\hbar^2  - \frac{4\nu^3+3\nu}{768}\hbar^3 -\frac{16\nu^4+72\nu^2+13}{49152}\hbar^4 +\cO(\hbar^5) \ ,
\end{equation}
with
\begin{equation}
	\nu = N + 1/2 \ .
\end{equation}
Indeed, this agrees with the perturbative energy of the Harper-Hofstadter model \eqref{eq:E-pert} by the replacement \eqref{eq:replace}.
Note we have adapted the series of $E^\text{pert}_{\bF_0}(\nu,\hbar)$ to be consistent with the normalization of  $\cH^{\bF_0}$ used in this paper. To formulate the results of the formal ``instanton--anti-instanton'' correction, we need some terminology from topological string theory on a local Calabi-Yau manifold and its mirror curve.

The coefficient $u$ in the equation of mirror curve \eqref{eq:F0-curve} parametrizes the complex structure moduli space of the curve. The moduli space has several singular points, one of which of particular interest is called the conifold singularity and it is located at $u=4$, as it corresponds to the semi-classical limit $E_{\bF_0}=0$ of the QM model $\cH_{\bF_0}$. Let us introduce 
\begin{equation}
	z = \frac{1}{u^2} \ .
\end{equation}
Then the classical periods of the mirror curve are
\begin{equation}
\begin{aligned}
	\pd_z t_c  &= -\frac{2}{\pi z} \md K(1-16z) \ ,\\
	\pd_z t_c^D &= \frac{2}{z\sqrt{1-16z}}\md K\(\frac{16}{16z-1}\) \ ,
\end{aligned}
\end{equation}
of which $t_c$ can serve as a good local coordinate on the moduli space near the conifold singularity. Here $\md K$ is the complete elliptic integral of the first kind. Furthermore, for the topological string theory on a local Calabi-Yau threefold $X$, an important quantity is the refined free energy $F(t,\epsilon_1,\epsilon_2)$. It encodes the numbers of BPS states of the M-theory compactified on $X\times (\bR^4\times S^1)_{\epsilon_1,\epsilon_2}$, where the parameters $\epsilon_1,\epsilon_2$ describe how the $\bR^4$ is twisted along $S^1$. $t$ is a set of coordinates on the moduli space of $X$, which due to mirror symmetry is mapped to the complex structure moduli space of the associated mirror curve. In the application to the spectrum of $\cH^{\bF_0}$, one is in particular interested in the so-called Nekrasov-Shatashvili limit \cite{nekrasov2010quantization}
\begin{equation}
	F^\text{NS}(t,\hbar) = \lim_{\epsilon_1\to 0}\ri\epsilon_1  F(t,\epsilon_1,\ri \hbar)\ ,
\end{equation}
and the free energy in the NS limit enjoys a genus expansion
\begin{equation}
	F^\text{NS}(t,\hbar) = \sum_{n=0}^{\infty} F_n^\text{NS}(t) \hbar^{2n} \ .
\end{equation}
Near the conifold singularity, the NS free energies $F_n^\text{NS}$ are functions of $t_c$ with at most logarithmic singularity, and we will use the notation
\begin{equation}
	F^\text{NS}(t,\hbar) = F^C(t_c,\hbar) = \sum_{n=0}^{\infty} F^C_n(t_c) \hbar^{2n} \ .
\end{equation}
They can be computed recursively by the so-called refined holomorphic anomaly equations \cite{Bershadsky:1993cx, Krefl:2010fm, Huang:2010kf} in the NS limit, as explained in detail in \cite{Codesido:2016dld,Codesido:2017jwp, Fischbach:2018yiu}. For local $\bF_0$, the first few NS free energies are
\begin{equation}
\begin{aligned}
	F_0^C(t_c) &= \frac{1}{2}t_c^2 \( \log \(\frac{t_c}{16}\) -\frac{3}{2}\) -\frac{t_c^3}{48} +\frac{5 t_c^4}{4608} -\frac{7 t_c^5}{61440} +\frac{733 t_c^6}{44236800} + \cO(t_c^7)  \ .\\
	F_1^C(t_c) &= -\frac{1}{24}\log t_c -\frac{11 t_c}{192} +\frac{49 t_c^2}{9216} -\frac{77 t_c^3}{73728} +\frac{2213 t_c^4}{8847360} -\frac{607 t_c^5}{9437184} + \cO(t_c^6)\ ,\\
	F_2^C(t_c) &= -\frac{7}{5760 t_c^2} -\frac{889 t_c}{2949120}+\frac{181981 t_c^2}{707788800} -\frac{16157 t_c^3}{113246208} +\frac{2194733 t_c^4}{32614907904} + \cO(t_c^5) \ .
\end{aligned}
\end{equation}
We stress that these results are obtained purely in the framework of topological string theory.
We do not need any knowledge of the corresponding quantum mechanics.
Our goal is to relate these quantities to the eigen-energy in quantum mechanics.

It turns out, the formal ``instanton--anti-instanton'' correction to the eigen-energy of $\cH^{\bF_0}$, which controls the asymptotic growth of the coefficients of $E^{\text{pert}}_{\bF_0}(\nu,\hbar)$, is given by \cite{Codesido:2017jwp}
\begin{equation}
	E^{I\bar{I}}_{\bF_0}(\nu,\hbar) = \pm\ri \,2f^{(1)} \re^{16\Catalan/\hbar} \frac{\pd E^\text{pert}_{\bF_0}(\nu,\hbar)}{\pd \nu} \exp\(-\frac{2}{\hbar}\frac{\pd F^C(t_c,\hbar)}{\pd t_c}\) \Big|_{t_c \to \hbar\nu} \ ,
\end{equation}
where $\Catalan$ is the Catalan's constant, and $f^{(1)}$ a free constant. 
The exponential factor is $\re^{16\Catalan/\hbar}=\re^{2A/\hbar}$, and this indeed corresponds to the 2-instanton sector in our terminology.
Using the NS free energies of local $\bF_0$, one can write down the terms in the exponential
\begin{equation}
\begin{aligned}
	-\frac{1}{\hbar}\frac{\pd F^C}{\pd t_c}\Big|_{t_c \to \hbar\nu} &= \nu - \nu \log \(\frac{\nu}{16}\) + \frac{1}{24\nu} - \frac{7}{2880\nu^3} + \cO(\nu^{-5}) \\
	& -\nu\log \hbar + \frac{12\nu^2+11}{192}\hbar -\frac{20\nu^3+49\nu}{4608}\hbar^2 + \frac{1680\nu^4+9240\nu^2 + 889}{2949120}\hbar^3 + \cO(\hbar^4) \ .
\end{aligned}
\end{equation}
Interestingly, the terms independent of $\hbar$ can be resummed to 
\begin{equation}
	\log\(\frac{\sqrt{2\pi}16^\nu}{\Gamma(\tfrac{1}{2}+\nu)}\) \ .
\end{equation}
Furthermore, let us denote the power series in $\hbar$ starting from $\cO(\hbar)$ by 
\begin{equation}
	\[-\frac{1}{\hbar}\frac{\pd F^C}{\pd t_c}\] \ .
\end{equation}
Then the ``instanton--anti-instanton'' correction can be written as
\begin{equation}\label{eq:E1-F0}
	E^{I\bar{I}}_{\bF_0}(\nu,\hbar)  = \pm\ri f^{(1)}\frac{ 2^{8\nu+2}\pi}{\Gamma(\tfrac{1}{2}+\nu)^2}\hbar^{1-2\nu} \re^{16\Catalan/\hbar} \cdot \frac{1}{\hbar}\frac{\pd E^\text{pert}_{\bF_0}(\nu,\hbar)}{\pd \nu}\exp\[-\frac{2}{\hbar}\frac{\pd F^C}{\pd t_c}\] \Big|_{t_c\to\hbar\nu} \ ,
\end{equation}
where the components after $\cdot$ is a power series starting from constant 1.

We observe here that this result in terms of topological string free energies also reproduces the imaginary part of the instanton--anti-instanton correction \eqref{eq:E-bion} for the Harper-Hofstadter model after applying the map \eqref{eq:replace}. Indeed, if we do so, we find that the factor in front of $\cdot$ in \eqref{eq:E1-F0} agrees with the prefactor before $\cdot$ in \eqref{eq:E-bion}, if we choose
\begin{equation}
	f^{(1)} =  \frac{1}{2\pi} \ .
\end{equation}
Note that this normalization constant can also be fixed through the path integral calculation in section~\ref{sc:int-anti}. Comparing the remaining part with the numerical result \eqref{eq:bion-fluc}, one conjectures then the A-function should be identified with the opposite of the derivative of the NS free energy for local $\bF_0$, i.e.
\begin{equation}\label{eq:A-F0}
	\cA(N,\phi) = \[+\frac{1}{\hbar}\frac{\pd F^C}{\pd t_c}\]\Big|_{\substack{\hbar\to-\phi\phantom{ss}\\t_c\to -\phi\nu}} \ .
\end{equation}
We follow the calculation in \cite{Codesido:2017jwp} of the NS free energies for local $\bF_0$ by solving the NS holomorphic anomaly equations and push it to a few orders higher than what is explicitly given in \cite{Codesido:2017jwp}. We find 
\begin{align}
	&\[+\frac{1}{\hbar}\frac{\pd F^C}{\pd t_c}\]\Big|_{\substack{\hbar\to-\phi\phantom{ss}\\t_c\to -\phi\nu}} = \(\frac{\nu ^2}{16}+\frac{11}{192}\) \phi
	+\(\frac{5 \nu ^3}{1152}+\frac{49 \nu }{4608}\) \phi ^2\nn
	&\phantom{xxxxxxxx}+\(\frac{7 \nu ^4}{12288}+\frac{77 \nu ^2}{24576}+\frac{889}{2949120}\) \phi^3
	+\(\frac{733 \nu ^5}{7372800}+\frac{2213 \nu ^3}{2211840}+\frac{181981 \nu }{353894400}\) \phi^4 \nn
	&\phantom{xxxxxxxx}+\(\frac{47 \nu ^6}{2359296}+\frac{3035 \nu ^4}{9437184}+\frac{16157 \nu^2}{37748736}+\frac{112573}{3170893824}\) \phi^5\nn
	&\phantom{xxxxxxxx}+\(\frac{35921 \nu ^7}{8323596288}+\frac{2443337 \nu ^5}{23781703680}+\frac{2194733 \nu ^3}{8153726976}+\frac{652008227 \nu}{7990652436480}\) \phi^6\nn
	&\phantom{xxxxxxxx}+\(\frac{83347 \nu ^8}{84557168640}+\frac{1183937 \nu^6}{36238786560}+\frac{42157069 \nu ^4}{289910292480}+\frac{427007447 \nu ^2}{4058744094720}\right.\nn
	&\phantom{xxxxxxxx}\left.+\frac{1910609149}{324699527577600}\) \phi^7 + \cO(\phi^8) \ ,
\end{align}
and it agrees completely with the A-function \eqref{eq:A-func-num} from the numerical fit.

Finally, since the power series in the 1-instanton sector is given by the A-function as shown in \eqref{eq:inst-fluc}, we claim that the 1-instanton sector can also be expressed in terms of the NS free energy of local $\bF_0$. In fact, by plugging in \eqref{eq:A-F0} and carefully ironing out the prefactor, we find
\begin{equation}
	E_{(\theta_x,\theta_y)}^\text{1-inst}(N,\phi) =  \frac{\cos\theta_x+\cos \theta_y}{\pi}\re^{-A/\phi} \frac{\pd E^\text{pert}(N)}{\pd N} \text{Im} \exp\(+\frac{1}{\phi} \frac{\pd F^C}{\pd t_c}\)\Big|_{\substack{\hbar\to -\phi\phantom{sssssssss}\\t_c \to - \phi (N+1/2)}} \ .
\end{equation}
Note that after mapping $\hbar \to -\phi$ the exponential becomes purely imaginary, and we take its imaginary value in the expression above.

\section{Conclusion and discussion}
\label{sc:conclusion}

In this paper, our goal is to understand the peculiar band structure of the energy spectrum of the Harper-Hofstadter model in the semi-classical limit. According to the general philosophy of resurgence, the energy levels should be written as trans-series summing over contributions from all saddle points coupled differently to the Bloch's angles, and in addition, the large order growth of the perturbative sector is controlled by the ambiguity of the energy, which receives leading order contributions from the instanton--anti-instanton sector.

We used various techniques to compute the trans-series energy levels. The perturbative series is computed very conveniently using the extended \texttt{BenderWu} package \cite{Sulejmanpasic:2016fwr, Gu:2017ppx}. The 1-loop contributions to the 1-, 2-instanton sectors, and the energy ambiguity are obtained by a path integral calculation, albeit restricted to the ground state level. Higher order corrections in the 1-instanton sector and in the ambiguity (imaginary contributions of the instanton--anti-instanton sector) are computed using refined topological string techniques in connection with the local $\bF_0$ geometry inspired by a similar work \cite{Codesido:2017jwp}. All these results can be checked against numerical results, which can be computed exactly when the magnetic flux is $2\pi$ times a rational number, and they all agree perfectly. This validates all our techniques. In the process, we find that the perturbative--non-perturbative relation\footnote{This relation was first proposed in \cite{alvarez2002-quartic,alvarez2004-doublewell,alvarez2000-cubic,alvarez2000-generic,alvarez2002anharmonic} and later rediscovered by \cite{Dunne2014}.} relating the perturbative sector and the 1-instanton sector is not satisfied\footnote{This should not be taken as indication that such a relation does not exist, but just that the form is different.}, which is not that surprising since the Schr\"{o}dinger equation of the Harper-Hofstadter model is a difference rather than a second-order differential equation. On the other hand there still exists a curious relation between the three sectors: perturbative, 1-instanton, instanton--anti-instanton.

Clearly there are many open questions. The most pressing one is how to understand better the nature of instantons, and in particular the $[I_xI_y]$ instanton configuration, the treatment of which is rather ad hoc in this work. Namely the inclusion of the saddles, such as instantons, is expected to be dictated by the Picard-Lefschetz theory, and requires a decomposition of the path-integral cycle into the Lefschetz thimbles. We have not rigorously checked whether instanton configurations we analyze are a part of this decomposition, but have argued that they should contribute on physical grounds. The case of $[I_xI_y]$ is particularly interesting, as it is an object without an analogue in simple 1D quantum mechanics. Na\"ive application of the BZJ prescription yields a result in perfect agreement with the numerics. On the other hand the BZJ prescription was interpreted as contributions from saddles at infinity \cite{Behtash:2018voa}. It would be desirable to understand this better in the case at hand. It would also be nice to have a path integral understanding of the higher order corrections computed using topological string techniques. Furthermore, another real world model, one that describes electrons on a triangular lattice, is revealed to be connected to the topological string theory, with the target space being the canonical bundle of the three-point blow-up of $\bP^2$ \cite{Hatsuda:2017zwn}. One can explore whether the similar analysis can be applied in that model as well.

\section*{Acknowledgement}

We would like to thank Amir-Kian Kashani-Poor, Marcos Mari\~no and Mithat \"Unsal for valuable discussions. YH and TS would like to thank the organizers of the RIMS-iTHEMS International Workshop on Resurgence Theory (2017) in Kobe where this work was initiated. TS would also like to thank Hartmut Wittig and the Institute of Nuclear Physics in Mainz for their hospitality, where a part of this work was conducted.
The work of YH is supported  Rikkyo University Special Fund for Research and by JSPS KAKENHI Grant Number 18K03657.
JG is supported by the European Research Council (Programme ``Ideas'' ERC-2012-AdG 320769 AdS-CFT-solvable).

\begin{appendix}
\section{Instanton solution}\label{app:1-instanton}

We would like to solve the instanton profile from its equations of motion
\begin{subequations}\label{eq:eom_app}
	\begin{align}
	& \ri \dot{x} - \sin y = 0 \ , \label{eq:eom1_app}\\
	& \ri \dot{y} + \sin x = 0 \ . \label{eq:eom2_app}
	\end{align}
\end{subequations}
We take the derivative w.r.t.\ time on \eqref{eq:eom1_app} and multiply it with $\dot{x}$, and after using \eqref{eq:eom2_app} to remove all appearance of $y$, we find
\begin{equation}
	\frac{\rd}{\rd t} (\sqrt{1+\dot{x}^2} \pm \cos x) = 0 \ ,
\end{equation}
where $\pm$ comes from converting $\cos y$ to $\sin y$, and the above equation integrates to the identity
\begin{equation}\label{eq:Econs1}
	E(\beta) = \sqrt{1+\dot{x}^2} \pm \cos x \ .
\end{equation}
We interpret the integration constant $E(\beta)$ to be the conserved energy of the saddle point configuration.  Indeed, when $\dot{x}$ is small, the r.h.s.\ of \eqref{eq:Econs1} becomes
\begin{equation}
	\frac{1}{2}\dot{x}^2 + 1 \pm \cos x
\end{equation}
which resembles the conserved energy of a saddle point configuration in non-relativistic QM where $1\pm \cos x$ is the inverted potential. For the 1-instanton configuration $x_1(t)$, the energy $E(\beta)$ reaches the maximum value in the limit $\beta\rightarrow \infty$, and it corresponds to the oscillation between two neighboring highest points of the inverted potential. In \eqref{eq:Econs1}, we have $E(\infty) = 2$ regardless of the sign in the inverted potential, so we simply take $+$ without loss of generality
\begin{equation}\label{eq:inst1-Econs}
	\sqrt{1+\dot{x_1}^2} + \cos x_1 = 2 \ .
\end{equation}
Solving \eqref{eq:inst1-Econs}, we find the following profile of 1-instanton
\begin{equation}\label{eq:inst1-x}
	x_1(t) = 2\cos^{-1} \( -\frac{\sqrt{2}\tanh(t-t_0)}{\sqrt{1+\tanh^2(t-t_0)}}\) \ ,
\end{equation}
as well as
\begin{equation}\label{eq:inst1-p}
	y_1(t) = \cos^{-1}(2-\cos(x_1(t))) = \cos^{-1} \( 1+\frac{2}{\cosh2(t-t_0)}\) \ .
\end{equation}

Using the conservation law 
\be
\cos y_1 + \cos x_1 = 2\ ,
\ee
we find that the action is given by
\begin{align}\label{eq:inst-action}
	A =& \int_{-\infty}^\infty \rd t (-\cos x_1 -\cos y_1+2 - \ri \dot{x}_1 y_1)  \nn
	=& -\ri \int_0^{2\pi} y_1(x_1) \rd x_1 = 2 \int_{0}^{\pi} \cosh^{-1}(2-\cos x) \rd x \nn
	=& 2\int_{0}^{\pi} \log\(2-\cos x +\sqrt{(3-\cos x)(1-\cos x)}\) \rd x \nn
	=& 4\int_{0}^{\pi} \log\( \sin \frac{x}{2} + \sqrt{1+\sin^2\frac{x}{2}}\)\rd x\nn
	=& 8\int_{0}^{1} \frac{\rd t}{1-t^2} \log(t+\sqrt{1+t^2}) = 8\Catalan \ .
\end{align}
In the last line we performed the change of variables $t =\sin x/2$, and used one of the definitions of the Catalan's constant
\begin{equation}
	\Catalan = \int_{0}^{1} \frac{\sinh^{-1}t}{\sqrt{1-t^2}} \rd t\  .
\end{equation}

\section{The moduli-space metric}\label{app:moduli-space metric}

We want to find the moduli-space metric of the one instanton. We can do this by adding a factor $\lambda \delta \tilde x^2$ to the action \eqref{eq:inst_inter} before integration, so as to lift the zero mode. Upon modifying \eqref{eq:inst_inter} by adding such a term, we can do the Gaussian integral and simply get
\be\label{eq:lambda_det}
Z_1^\lambda=\frac{1}{\sqrt{\stO+\lambda}}\;.
\ee

Now let us write the small deviation around the instanton solution as
\be\label{eq:tildex_decomp}
\delta \tilde x\approx \({\partial_{t_0} x_1}\big|_{t_0=0} /\sqrt{\cos y_1}\)t_0 +\delta \tilde x^\perp= -\dot x_1 t_0/\sqrt{\cos y_1}+\delta \tilde x^\perp\;,
\ee
where $\delta \tilde x$ is orthogonal to $\cos y_1 \dot x_1$. The first term is a small deviation from the instanton solution in the direction of the zero mode, and $t_0$ specifies a shift of its position in time. In fact $t_0$ is precisely the coordinate we want to isolate, and over which we will integrate exactly, producing a factor of $\beta$. Recall that our goal is to find a way to write the path integral in \eqref{eq:inst_inter}, as 
\be
Z_1=\int \rd t_0\;  \; \frac{\mu}{\sqrt{\det' \stO}}\ ,
\ee
where the prime indicates that the zero-mode has been excluded from the determinant. The $\mu$ above is the measure of the zero-mode moduli $t_0$ (also referred to as moduli space metric), which is what we wish to find.

To find it we will add the term $\lambda \delta\tilde x^2$ into the action as before, and integrate over $t_0$. We should get \eqref{eq:lambda_det}, up to a constant, which will precisely correspond to $\mu^{-1}$. To do this, let us plug in the expression \eqref{eq:tildex_decomp} for $\delta \tilde x$ into the path integral \eqref{eq:inst_inter}. It only amounts to adding the term $\lambda \delta \tilde x^2$ into the action, since the zero mode is annihilated by $\stO$. Then it is easy to see that the action contains the term
\be
	e^{-\frac{\lambda N^2}{2\phi}t_0^2}\;.
\ee
where N is given by \eqref{eq:zeromode_norm}.
If we now integrate over $t_0$ and $\delta \tilde x^\perp$ we produce a term
\be
	\sqrt{\frac{2\pi \phi }{\lambda N^2}} \frac{\mu}{\sqrt{\det' (\stO+\lambda)}}\ ,
\ee 
where the prime on the determinant means we have excluded the zero mode of the $\stO$ operator. The $\lambda$ in the denominator however combines with the primed determinant to give the complete determinant
\be
	\sqrt{\frac{2\pi \phi }{N^2}} \frac{\mu}{\sqrt{\det(\stO+\lambda)}}\;.
\ee 
Comparing with \eqref{eq:lambda_det}, we can read off the measure to be
\be
	\mu=\sqrt{\frac{N^2}{2\pi \phi}}\ .
\ee

\section{The one-instanton determinant}\label{app:determinant}

We will compute the determinant of the one-instanton fluctuation operator using the Gel'fand-Yaglom theorem, explained for instance in \cite{coleman1988aspects,Dunne:2007rt,Takhtajan:2008,Marino:2015yie}. Consider an ordinary differential operator $\sO$, with a canonical second derivative term $\sO=-\partial_t^2+\dots$. We wish to compute  the determinant of the operator. For that purpose we consider the space of functions on which the operator acts  to be defined on an interval $t\in [-\beta/2,\beta/2]$ with the Dirichlet boundary conditions\footnote{More appropriate boundary conditions for computing path-integral determinants would be periodic boundary conditions, as Euclidean time is periodic. However in the limit of large Euclidean time-expanse -- the limit relevant for the ground state properties of the system -- the boundary conditions do not matter. Since the formulas are simpler when Dirichlet boundary conditions are used. But everything can be generalized to periodic boundary conditions if so desired. Indeed if one wished to study the excited spectrum of the theory, one would need to do precisely this. } for the eigenfunctions $\phi(t)$, i.e.
\begin{equation}\label{eq:bc-1}
	\phi(-\beta/2) = \phi(\beta/2) = 0 \ .
\end{equation}
Then the Gel'fand-Yaglom theorem states that the determinant of the operator $\sO$ is
\begin{equation}\label{eq:GY-1}
\det \sO \propto \Psi(\beta/2)  \ ,
\end{equation}
where $\Psi(t)$ is a zero mode of $\sO$, i.e.
\begin{equation}
	\sO \circ \Psi(t) = 0
\end{equation}
satisfying a \emph{different} boundary condition
\begin{equation}\label{eq:bc-2}
	\Psi(-\beta/2) = 0 \ ,\quad \dot{\Psi}(-\beta/2) = 1 \ .
\end{equation}
The proportionality identity can be made precise by regularizing the operator determinant with that of a simple operator, for instance, the harmonic oscillator
\begin{equation}\label{eq:GY-2}
	\frac{\det \sO}{\det \sO_0} = \frac{\Psi(\beta/2)}{\Psi_0(\beta/2)} \ ,
\end{equation}
where $\Psi_0(t)$ is the zero mode of the harmonic oscillator $\sO = -\partial_t^2 + 1$ with the boundary condition \eqref{eq:bc-2}, and it is simply
\begin{equation}\label{eq:Psi-h}
	\Psi_0(t) = \sinh(t+\beta/2) \ .
\end{equation}\label{eq:GY-3}
To treat $\det' \sO$ with zero mode removed, we can use the relation
\begin{equation}\label{eq:detpO}
	\det\nolimits' \sO = \lim_{\lambda \rightarrow 0} \frac{\rd}{\rd \lambda} \det\sO_\lambda \ ,
\end{equation}
with 
\begin{equation}
	\sO_\lambda := \sO + \lambda \ .
\end{equation}
Therefore we need to compute the zero mode of $\sO_\lambda$ satisfying the boundary condition \eqref{eq:bc-2} up to order $\lambda$.

Now we could take the operator $\sO$ to simply be the fluctuation operator $\stO$ given by \eqref{eq:Otilda}. However notice that we have 
\be
\det(\stO+\lambda)=\det[f(t)\stO \frac{1}{f(t)}+\lambda]\;,
\ee
where $f(t)$ is an arbitrary, nonsingular function with no zeros. By taking the derivative with respect to $\lambda$ and setting $\lambda=0$ we get
\be
\det{}'(\stO)=\det{}'[f(t)\stO \frac{1}{f(t)}]\ .
\ee

If we take $f(t)=\sqrt{\cos y_1(t)}$ we can define the operator 
\be
\sO={\sqrt{\cos y_1(t)}}\stO\frac{1}{\sqrt{\cos y_1(t)}}=\cos y_1\left(-\pd_t \frac{1}{\cos y_1(t)}\pd_t + \cos x_1(t)\right)
\ee
so that we will compute $\det{}'(O)$ instead of $\det{}(O)$.

In order to compute it we first have to consider the determinant of $\det \sO_\lambda$, where $\sO_\lambda=\sO+\lambda$, at least for small $\lambda$. We already know that $\sO$ has a zero mode given by $\dot x_1$.  To use Gel'fand-Yaglom theorem we look for a solution
\begin{equation}\label{eq:Olamb}
	\sO_\lambda \Psi_\lambda = 0 \ ,
\end{equation}
where $\Psi_\lambda$ satisfies \eqref{eq:bc-2}. Now assuming $\lambda$ is small we can write
\begin{equation}
	\Psi_\lambda  = \Psi^{(0)} + \lambda \Psi^{(1)}(t) + \cO(\lambda^2) \ ,
\end{equation}
where
\be
\sO\Psi^{(0)}=0
\ee
and
\be\label{eq:second-order}
\sO\Psi^{(1)}=-\Psi^{(0)}\;.
\ee
The first of these equations reduces to
\begin{equation}\label{eq:O0lamb}
	\sO \Psi^{(0)} ={\cos y_1(t)} \(-\pd_t \frac{1}{\cos y_1(t)}\pd_t + \cos x_1(t)\) \Psi^{(0)} = 0 \ .
\end{equation}
This is a second order ODE, and we already know that one solution is 
\begin{equation}\label{eq:psi-1}
	\psi_1(t) = \dot{x}_1(t)\ ,
\end{equation}
although it does not satisfy the boundary condition \eqref{eq:bc-2}. In order to find a second independent solution, we notice that the operator $\sO$ can be factorized in the following way. We introduce operators
\begin{equation}
	\sQ = \frac{1}{\cos y_1} \pd_t - \ri\, \frac{\sin x_1}{\sin y_1} \ ,\quad \sQ^\dagger = \frac{1}{\cos y_1} \pd_t + \ri\, \frac{\sin x_1}{\sin y_1} \ .
\end{equation}
They satisfy
\begin{equation}\label{eq:QQ}
	\sQ^\dagger \sQ= -\frac{1}{\cos^2 y_1}\sO \ ,\quad \sQ \sQ^\dagger = -\frac{1}{\cos^2 y_1} \sO + \frac{2}{\cos y_1}\(\frac{\cos x_1}{\cos y_1}+\frac{\sin^2 x_1}{\sin^2 y_1}\) \ .
\end{equation}
We want to find the most general homogeneous solution to the equation $\sO \psi=0$. This is the same as finding such a solution for the operator $\sQ^\dagger \sQ$. We observe that $\sQ^\dagger$ annihilates $1/\dot{x}_1$. If one can find $\psi_2$ such that $\sQ \psi_2 = 1/\dot{x}_1$, then one concludes immediately from \eqref{eq:QQ} that $\psi_2$ is another solution to \eqref{eq:O0lamb}. Indeed by making an appropriate ansatz we find
\begin{equation}\label{eq:psi-2}
	\psi_2(t) = \dot{x}_1(t)\int^{t} \rd t' \frac{\cos y_1(t')}{\dot{x}_1^2(t')} \ .
\end{equation}
Furthermore since the Wronskian is not identically vanishing
\begin{equation}\label{eq:wron}
	W_{21}(t) := \psi_2(t) \pd_t \psi_1(t) - \psi_1(t)\pd_t \psi_2(t) = -\cos y_1(t)\ ,
\end{equation}
the two solutions are linearly independent. From $\psi_2(t)$ we can construct the solution to \eqref{eq:O0lamb} satisfying the boundary condition \eqref{eq:bc-2}
\begin{equation}
	\Psi^{(0)}(t) = \frac{\dot{x}_1(-\beta/2)}{\cos y_1(-\beta/2)} \dot{x}_1(t) \int_{-\beta/2}^{t}\rd t' \frac{\cos y_1(t')}{\dot{x}_1^2(t')} \ .
\end{equation}

Let us proceed to the next order in $\lambda$, namely Eq.~\eqref{eq:second-order},
\begin{equation}\label{eq:O1lamb}
	\(\pd_t^2 - \frac{\dot{y}_1\sin y_1 }{\cos y_1}\pd_t - \cos y_1\cos x_1\) \Psi^{(1)} = -\sO\circ \Psi^{(1)} = \Psi^{(0)} \ ,
\end{equation}
and $\Psi^{(1)}(t)$ satisfies the boundary condition
\begin{equation}\label{eq:bc-3}
	\Psi^{(1)}(-\beta/2) = 0 \ ,\quad \dot{\Psi}^{(1)}(-\beta/2) = 0 \ .
\end{equation}
One way to solve \eqref{eq:O1lamb} is to first find the modified Green's function $G(t,t')$ satisfying
\begin{equation}\label{eq:greeneq}
	 \sO G(t,t') = \cos y_1 \delta (t-t') \ ,
\end{equation}
so that $\Psi^{(1)}$ is given by
\begin{equation}
	\Psi^{(1)}(t) = \int_{-\beta/2}^{\beta/2} \rd t' G(t,t') \Psi^{(0)}(t')\frac{1}{\cos y_1} \ .
\end{equation}
We claim that the Green's function is given by
\begin{equation}\label{eq:greenfunc}
	G(t,t') = \begin{cases}
	-\psi_1(t)\psi_2(t') + \psi_2(t)\psi_1(t') \ ,&t> t' \ ,\\
	0 \ , & t\leq t' \ .
\end{cases}
\end{equation}
Indeed, when both $t<t'$ and $t>t'$, \eqref{eq:greeneq} is trivially satisfied since $\psi_1(t), \psi_2(t)$ are annihilated by $\sO$. In the neighborhood of $t\rightarrow t'$, let us plug \eqref{eq:greenfunc} into \eqref{eq:greeneq}, integrate both sides from $ t= t'-\epsilon$ to $t=t'+\epsilon$ and take the limit $\epsilon\rightarrow 0$. The r.h.s.\ is simply $\cos y_1(t')$, while the l.h.s.\ is given by
\begin{equation}
	\pd_t G(t,t')\big|_{t=t'^{+}} - \pd_t G(t,t')\big|_{t=t'^{-}} = 
	-W_{21}(t') = \cos y_1(t') \ ,
\end{equation}
where we have used \eqref{eq:wron}. Therefore \eqref{eq:greenfunc} is the correct modified Green's function. We can now write down $\Psi^{(1)}(t)$
\begin{equation}\label{eq:Psi-1}
	\Psi^{(1)}(t) =  \int_{-\beta/2}^{t}\rd t' \Psi^{(0)}(t')\frac{1}{\cos y_1(t')}\(\psi_1(t')\psi_2(t) - \psi_2(t')\psi_1(t)\) \ .
\end{equation}
This function indeed satisfies the boundary condition \eqref{eq:bc-3}.

Now we are ready to compute the operator determinant using the Gel'fand-Yaglom theorem. Combining \eqref{eq:GY-2},\eqref{eq:Psi-h},\eqref{eq:detpO},\eqref{eq:Psi-1}, we have
\begin{equation}
	\frac{\det\nolimits'\sO}{\det \sO_0} = \frac{\det\nolimits'\stO}{\det \sO_0}=\frac{\dot{x}_1(-\beta/2)\dot{x}_1(\beta/2)}{\sinh \beta\cos y_1(-\beta/2)}\int_{-\beta/2}^{\beta/2}\rd t\, \frac{\dot{x}^2_1(t)}{\cos y_1} \int_{-\beta/2}^{t}\rd t' \frac{\cos y_1(t')}{\dot{x}^2_1(t')}\int^{\beta/2}_{t}\rd t'' \frac{\cos y_1(t'')}{\dot{x}^2_1(t'')} \ .
\end{equation}

\section{Comparison of Fourier analysis with linear formulas }\label{app:fourier}

In the section~\ref{sc:n2inst} we have used \eqref{eq:2inst-lin} to compute the 2-instanton $\theta_x,\theta_y$-dependence of the system. These formulas are expected to have an error exponentially suppressed with the coupling. Here we analyze  $E_0^\text{pert} + E_0^{I\bar{I}}, E_0^{\text{1-inst}}, E_0^{II}, E_0^{II\text{mix}}$ by a direct Fourier transformation in order to check explicitly the validity of the formulas \eqref{eq:2inst-lin} and make an estimate on the error. To be explicit, we should have
\begin{equation}\label{eq:lin-combs}
\begin{aligned}
	E_0^\text{pert} + E_0^{I\bar{I}} &= \frac{1}{\pi^2}\int_{0}^{\pi}\rd \theta_x\rd\theta_yE_0(\theta_x,\theta_y)\ , \\
	E_0^{\text{1-inst}} &= \frac{2}{\pi}\int_{0}^{\pi}\rd \theta_x \cos(\theta_x) E_0(\theta_x,\tfrac{\pi}{2}) \ , \\
	E_0^{II} &= \frac{2}{\pi}\int_{0}^{\pi}\rd \theta_x \cos(2\theta_x) E_0(\theta_x,\tfrac{\pi}{2})\ , \\
	E_0^{II\text{mix}} &= \frac{4}{\pi^2}\int_{0}^{\pi}\rd \theta_x\rd\theta_y \cos(\theta_x) \cos(\theta_y) E_0(\theta_x,\theta_y)\ . \\
\end{aligned}
\end{equation}
Here we can integrate over $\theta \in [0,\pi]$ instead of $[0,2\pi]$ because $E_0$ is an even function of $\theta_x,\theta_y$.

We notice that the energy level $E_0$ is solved from the equation (c.f.\ \eqref{eq:Fab-simp})
\begin{equation}\label{eq:FE}
	F_Q(E_0) = 2(\cos\theta_x + \cos \theta_y -2)
\end{equation}
where $F_Q(E_0)$ for $\phi= 2\pi/Q$ is a polynomial of degree Q in $E_0$. It is simpler to perform the integration if we can exchange the integration variable from $\theta_x,\theta_y$ to $E_0$. This can be done in the following way.

To compute $E_0^{\text{1-inst}}$, we take $\theta_y = \pi/2$ and $E_0$ only depends on $\cos\theta_x = \cos\theta$. The relation \eqref{eq:FE} is simplified to
\begin{equation}\
	F_Q(E_0) = 2(\cos\theta-2) \ .
\end{equation}
Then we shall have
\begin{align}
	E_0^{\text{1-inst}} &= \frac{2}{\pi}\int_{0}^{\pi}\rd \theta \cos(\theta) E_0(\cos\theta) \nn
	&= \frac{2}{\pi}\int_{-1}^{+1}\rd \cos \theta \frac{\cos\theta}{\sqrt{1-\cos^2\theta}} E_0(\cos\theta) \nn
	&= \frac{2}{\pi}\int_{E_0(-1)}^{E_0(1)}\rd E_0 E_0 \frac{\rd \cos \theta}{\rd E_0} \frac{\cos \theta}{\sqrt{1-\cos^2\theta}} \nn
	&=\frac{2}{\pi}\int_{E_0(-1)}^{E_0(1)}\rd E_0 E_0 \frac{1}{2}F'_Q(E_0)\frac{\frac{1}{2}F_Q(E_0)+2}{\sqrt{1-(\frac{1}{2}F_Q(E_0)+2)^2}} \ . \label{eq:Ix}
\end{align}

Similarly, we have for the $E_0^{II}$ 
\begin{align}
	E_0^{II} &= \frac{2}{\pi} \int_0^\pi \rd \theta \cos(2\theta)E_0(\cos\theta) \nn 
	&= \frac{2}{\pi} \int_{-1}^1 \rd \cos\theta \frac{2\cos^2\theta-1}{\sqrt{1-\cos^2\theta}} E_0(\cos\theta) \nn
	&=\frac{2}{\pi} \int_{E_0(-1)}^{E_0(1)} \rd E_0 E_0 \frac{1}{2} F'_Q(E_0) \frac{2(\frac{1}{2}F_Q(E_0)+2)^2-1}{\sqrt{1-(\frac{1}{2}F_Q(E_0)+2)^2}}   \ . \label{eq:Ixx}
\end{align}

In the case of $E_0^\text{pert}+E_0^{I\bar I}$ and $E_0^{II\text{mix}}$, the energy $E_0$ depends on $\cos\theta_x + \cos\theta_y$. Let us define
\begin{equation}\
\begin{aligned}
	s = \cos\theta_x + \cos\theta_y \ ,\quad t = \cos\theta_x - \cos\theta_y \ .
\end{aligned}
\end{equation}
The integration range $\theta_x \in [0,\pi]$, $\theta_y \in [0,\pi]$ is equivalent to
\begin{equation}
	s\in [-2,2]
\end{equation}
and
\begin{equation}\label{eq:t-range}
	t\in \begin{cases}
		[-s-2,s+2] & s<0 \\
		[s-2,-s+2] & s>0 
	\end{cases} \ .
\end{equation}
We find that $E_0^\text{pert}+E_0^{I\bar I}$ is computed by
\begin{align}
	E_0^\text{pert}+E_0^{I\bar I} &= \frac{1}{\pi^2} \int_{0}^{\pi} \rd \theta_x \rd \theta_y E_0(\cos \theta_x+\cos\theta_y) \nn 
	&= \frac{1}{\pi^2} \int_{-1}^1 \frac{\rd \cos \theta_x \rd \cos \theta_y}{\sqrt{(1-\cos^2\theta_x)(1-\cos^2\theta_y)}} E_0(\cos \theta_x+\cos\theta_y) \nn
	&=\frac{1}{2\pi^2} \int \frac{\rd s\rd t}{\sqrt{1-\frac{s^2+t^2}{2}+ (\frac{s^2-t^2}{4})^2}} E_0(s) \nn
	&=\frac{1}{\pi^2} \int_{-2}^{2} \rd s E_0(s) \int_0^{\pm s+2} \frac{4\rd t}{\sqrt{16-8(s^2+t^2)+(s^2-t^2)^2}} 
\end{align}
where the integration range for $t$ is $[0,s+2]$ if $s<0$ and $[0,-s+2]$ if $s>0$.
The integration on $t$ can be performed explicitly, and we find
\begin{equation}
	\mathcal{K}(s):=\int_0^{\pm s+2} \frac{4 \rd t}{\sqrt{16-8(s^2+t^2)+(s^2-t^2)^2}} = \begin{cases}
		\frac{4}{2-s}{\mathbf K}\[\( \frac{2+s}{2-s}\)^2\]  & s<0 \\
		\frac{4}{2+s}{\mathbf K}\[\(\frac{2-s}{2+s}\)^2\] & s>0
	\end{cases} \ ,
\end{equation}
where ${\mathbf K}(\bullet)$ is the complete elliptic integral of the first kind. In the end, we have
\begin{equation}\label{eq:Ixxbar}
	E_0^\text{pert}+E_0^{I\bar I} = \frac{1}{2\pi^2}\int_{E_0(-2)}^{E_0(2)} \rd E_0 E_0 F'_Q(E_0) \mathcal{K}\(\tfrac{1}{2}F_Q(E_0)+2\) \ .
\end{equation}

Similarly for $E_0^{II\text{mix}}$, we have
\begin{align}\
	E_0^{II\text{mix}} &= \frac{4}{\pi^2} \int_{0}^\pi \rd \theta_x\rd\theta_y \cos \theta_x \cos \theta_y E_0(\cos\theta_x+\cos\theta_y) \nn
	&= \frac{4}{\pi^2} \int_{-1}^{1} \rd \cos\theta_x\rd \cos \theta_y \frac{\cos\theta_x\cos \theta_y}{\sqrt{(1-\cos^2 \theta_x)(1-\cos^2 \theta_y)}} E_0(\cos\theta_x+\cos\theta_y) \nn
	&=\frac{4}{\pi^2} \int_{-2}^{2} \rd s E_0(s) \int_0^{\pm s+2} \rd t \frac{s^2 -t^2}{\sqrt{16-8(s^2+t^2)+(s^2-t^2)^2}}  
\end{align}
We carry out the integration on $t$ explicitly
\begin{equation}\
	\mathcal{L}(s):=\int_0^{\pm s+2} \rd t \frac{s^2 -t^2}{\sqrt{16-8(s^2+t^2)+(s^2-t^2)^2}} =\begin{cases}
		(2-s)\mathbf E\[\(\frac{2+s}{2-s}\)^2\] + \frac{4(-1-s)}{2-s}\mathbf K\[\(\frac{2+s}{2-s}\)^2\] & s<0 \\
		(2+s)\mathbf E\[\(\frac{2-s}{2+s}\)^2\] + \frac{4(-1+s)}{2+s}\mathbf K\[\(\frac{2-s}{2+s}\)^2\] & s>0
	\end{cases} 
\end{equation}
where $\mathbf E(\bullet)$ is the complete elliptic integral of the second kind,
and conclude in the end
\begin{equation}\label{eq:Ixy}
	E_0^{II\text{mix}} = \frac{2}{\pi^2}\int_{E_0(-2)}^{E_0(2)} \rd E_0 E_0 F'_Q(E_0) \mathcal{L}\(\tfrac{1}{2}F_Q(E_0)+2\) \ .
\end{equation}
Equations \eqref{eq:Ixxbar},\eqref{eq:Ix},\eqref{eq:Ixx},\eqref{eq:Ixy} then provide us the means to compute the instanton contributions as Fourier coefficients of the ground state energy. 

\begin{figure}
	\centering
	\includegraphics[width=0.5\linewidth]{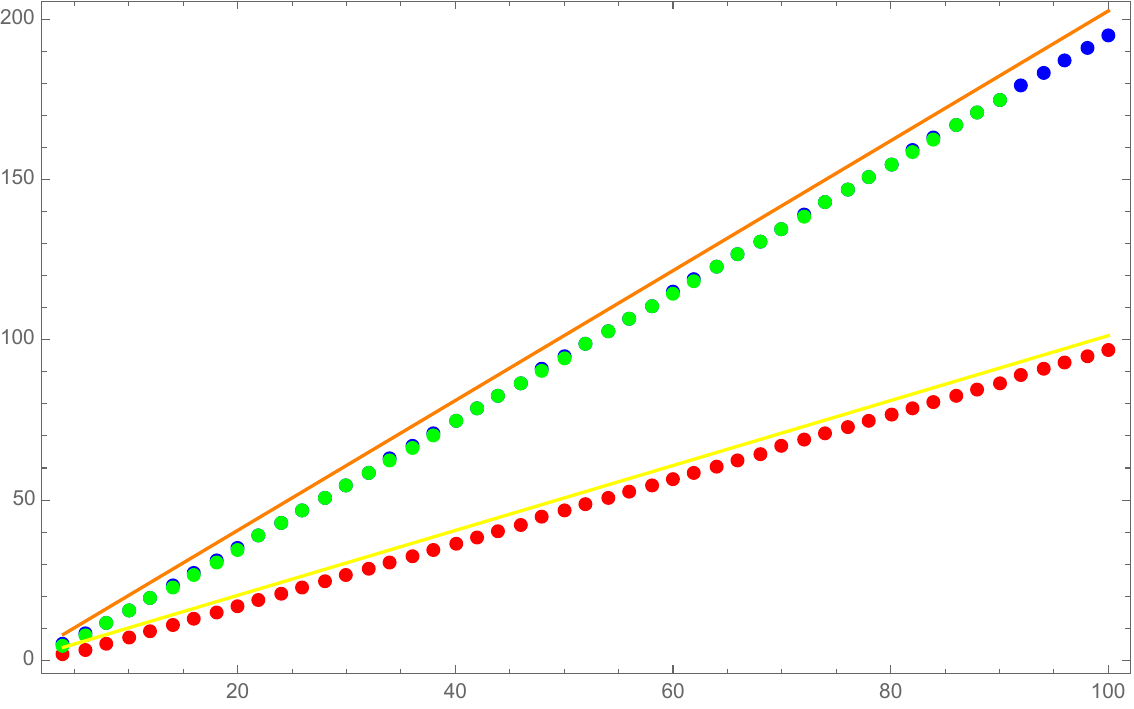}
	\caption{Plot of matching digits between results from linear combination and from Fourier transformation against $Q = 2\pi/\phi$ for various 2-instanton events. Blue, red, and green (indistinguishable from blue) dots correspond to $E_0^{II}, E_0^{II\text{mix}}$, and $E_0^\text{pert}+E_0^{I\bar{I}}$ respectively. Also plotted in yellow (orange) line is $\log_{10}$ of 2-(4-)instanton action $2A/\phi$ ($4A/\phi$).}\label{fg:fourier-linear}
\end{figure}

We compare the results of different 2-instanton corrections computed by the two different methods in Fig.~\ref{fg:fourier-linear}. Since we focus on the 2-instanton sector, we only include the corrections of $E_0^\text{pert}+E_0^{I\bar I}, E_0^{II}$, and $E_0^{II\text{mix}}$. We find the agreement to be remarkable, with the relative difference to be at 2-instanton or even 4-instanton levels, and is thus negligible.

\end{appendix}

\bibliographystyle{amsmod}
\bibliography{Hofstadter}

\ifx\undefined\bysame
\newcommand{\bysame}{\leavevmode\hbox to3em{\hrulefill}\,}
\fi
\begin{thebibliography}{10}

\bibitem{Harper:1955}
P.~G. Harper, {\em Single band motion of conduction electrons in a uniform
  magnetic field}, Proceedings of the Physical Society. Section A {\bf 68}
  (1955) 874.

\bibitem{Hofstadter:1976zz}
D.~R. Hofstadter, {\em {Energy levels and wave functions of Bloch electrons in
  rational and irrational magnetic fields}}, Phys. Rev. {\bf B14} (1976)
  2239--2249.

\bibitem{PhysRevB.41.11328}
D.~Freed and J.~A. Harvey, {\em Instantons and the spectrum of bloch electrons
  in a magnetic field}, Phys. Rev. B {\bf 41} (1990) 11328--11345.

\bibitem{Wilkinson305}
M.~Wilkinson, {\em Critical properties of electron eigenstates in
  incommensurate systems}, Proceedings of the Royal Society of London. Series
  A, Mathematical and Physical Sciences {\bf 391} (1984) 305--350.

\bibitem{Hatsuda:2016mdw}
Y.~Hatsuda, H.~Katsura, and Y.~Tachikawa, {\em {Hofstadter's butterfly in
  quantum geometry}}, New J. Phys. {\bf 18} (2016) 103023, {\tt
  arXiv:1606.01894} {\tt [hep-th]}.

\bibitem{Grassi:2014zfa}
A.~Grassi, Y.~Hatsuda, and M.~Marino, {\em {Topological strings from quantum
  mechanics}}, Annales Henri Poincare {\bf 17} (2016) 3177--3235, {\tt
  arXiv:1410.3382} {\tt [hep-th]}.

\bibitem{Hatsuda:2017zwn}
Y.~Hatsuda, Y.~Sugimoto, and Z.~Xu, {\em {Calabi-Yau geometry and electrons on
  2d lattices}}, Phys. Rev. {\bf D95} (2017) 086004, {\tt arXiv:1701.01561}
  {\tt [hep-th]}.

\bibitem{Sulejmanpasic:2016fwr}
T.~Sulejmanpasic and M.~\"{U}nsal, {\em {Aspects of perturbation theory in
  quantum mechanics: The BenderWu Mathematica package}}, {\tt arXiv:1608.08256}
  {\tt [hep-th]}.

\bibitem{Gu:2017ppx}
J.~Gu and T.~Sulejmanpasic, {\em {High order perturbation theory for difference
  equations and Borel summability of quantum mirror curves}}, JHEP {\bf 12}
  (2017) 014, {\tt arXiv:1709.00854} {\tt [hep-th]}.

\bibitem{Bershadsky:1993cx}
M.~Bershadsky, S.~Cecotti, H.~Ooguri, and C.~Vafa, {\em {Kodaira-Spencer theory
  of gravity and exact results for quantum string amplitudes}}, Commun. Math.
  Phys. {\bf 165} (1994) 311--428, {\tt arXiv:hep-th/9309140} {\tt [hep-th]}.

\bibitem{Krefl:2010fm}
D.~Krefl and J.~Walcher, {\em {Extended Holomorphic Anomaly in Gauge Theory}},
  Lett. Math. Phys. {\bf 95} (2011) 67--88, {\tt arXiv:1007.0263} {\tt
  [hep-th]}.

\bibitem{Huang:2010kf}
M.-x. Huang and A.~Klemm, {\em {Direct integration for general $\Omega$
  backgrounds}}, Adv. Theor. Math. Phys. {\bf 16} (2012) 805--849, {\tt
  arXiv:1009.1126} {\tt [hep-th]}.

\bibitem{Codesido:2017jwp}
S.~Codesido, M.~Mari\~{n}o, and R.~Schiappa, {\em {Non-perturbative quantum
  mechanics from non-perturbative strings}}, {\tt arXiv:1712.02603} {\tt
  [hep-th]}.

\bibitem{Codesido:2016dld}
S.~Codesido and M.~Mari\~{n}o, {\em {Holomorphic anomaly and quantum
  mechanics}}, {\tt arXiv:1612.07687} {\tt [hep-th]}.

\bibitem{Fischbach:2018yiu}
F.~Fischbach, A.~Klemm, and C.~Nega, {\em {WKB Method and Quantum Periods
  beyond Genus One}}, {\tt arXiv:1803.11222} {\tt [hep-th]}.

\bibitem{Hasegawa:1990}
Y.~Hasegawa, Y.~Hatsugai, M.~Kohmoto, and G.~Montambaux, {\em {Stabilization of
  flux states on two-dimensional lattices}}, phys. rev. B {\bf 41} (1990) 9174.

\bibitem{0305-4470-24-10-019}
J.~Bellissard, C.~Kreft, and R.~Seiler, {\em Analysis of the spectrum of a
  particle on a triangular lattice with two magnetic fluxes by algebraic and
  numerical methods}, Journal of Physics A: Mathematical and General {\bf 24}
  (1991) 2329.

\bibitem{10.21468/SciPostPhys.4.5.024}
J.~N. Fuchs, F.~Piéchon, and G.~Montambaux, {\em Landau levels, response
  functions and magnetic oscillations from a generalized onsager relation},
  SciPost Phys. {\bf 4} (2018) 24.

\bibitem{Witten:2010zr}
E.~Witten, {\em {A new look at the path integral of quantum mechanics}}, {\tt
  arXiv:1009.6032} {\tt [hep-th]}.

\bibitem{Witten:2010cx}
E.~Witten, {\em {Analytic Continuation Of Chern-Simons Theory}}, AMS/IP Stud.
  Adv. Math. {\bf 50} (2011) 347--446, {\tt arXiv:1001.2933} {\tt [hep-th]}.

\bibitem{Harlow:2011ny}
D.~Harlow, J.~Maltz, and E.~Witten, {\em {Analytic continuation of Liouville
  theory}}, JHEP {\bf 12} (2011) 071, {\tt arXiv:1108.4417} {\tt [hep-th]}.

\bibitem{Behtash:2015kna}
A.~Behtash, T.~Sulejmanpasic, T.~Sch�fer, and M.~�nsal, {\em {Hidden
  topological angles and Lefschetz thimbles}}, Phys. Rev. Lett. {\bf 115}
  (2015) 041601, {\tt arXiv:1502.06624} {\tt [hep-th]}.

\bibitem{Behtash:2015kva}
A.~Behtash, E.~Poppitz, T.~Sulejmanpasic, and M.~�nsal, {\em {The curious
  incident of multi-instantons and the necessity of Lefschetz thimbles}}, JHEP
  {\bf 11} (2015) 175, {\tt arXiv:1507.04063} {\tt [hep-th]}.

\bibitem{Behtash:2015zha}
A.~Behtash, G.~V. Dunne, T.~Sch\"afer, T.~Sulejmanpasic, and M.~\"Unsal, {\em
  {Complexified path integrals, exact saddles and supersymmetry}}, Phys. Rev.
  Lett. {\bf 116} (2016) 011601, {\tt arXiv:1510.00978} {\tt [hep-th]}.

\bibitem{Behtash:2015loa}
A.~Behtash, G.~V. Dunne, T.~Sch\"afer, T.~Sulejmanpasic, and M.~\"Unsal, {\em
  {Toward Picard-Lefschetz theory of path integrals, complex saddles and
  resurgence}}, {\tt arXiv:1510.03435} {\tt [hep-th]}.

\bibitem{Kozcaz:2016wvy}
C.~Kozcaz, T.~Sulejmanpasic, Y.~Tanizaki, and M.~\"Unsal, {\em {Cheshire Cat
  resurgence, self-resurgence and quasi-exact solvable systems}}, {\tt
  arXiv:1609.06198} {\tt [hep-th]}.

\bibitem{Behtash:2018voa}
A.~Behtash, G.~V. Dunne, T.~Sch\"afer, T.~Sulejmanpasic, and M.~\"Unsal, {\em
  {Critical points at infinity, non-Gaussian saddles, and bions}}, {\tt
  arXiv:1803.11533} {\tt [hep-th]}.

\bibitem{Nekrasov:2018pqq}
N.~Nekrasov, {\em {Tying up instantons with anti-instantons}}, 2018,
  pp.~351--388, {\tt arXiv:1802.04202} {\tt [hep-th]}.

\bibitem{Hatsuda:2017dwx}
Y.~Hatsuda, {\em {Perturbative/nonperturbative aspects of Bloch electrons in a
  honeycomb lattice}}, {\tt arXiv:1712.04012} {\tt [hep-th]}.

\bibitem{BOGOMOLNY1980431}
E.~Bogomolny, {\em Calculation of instanton-anti-instanton contributions in
  quantum mechanics}, Physics Letters B {\bf 91} (1980) 431 -- 435.

\bibitem{ZINNJUSTIN1981125}
J.~Zinn-Justin, {\em Multi-instanton contributions in quantum mechanics},
  Nuclear Physics B {\bf 192} (1981) 125 -- 140.

\bibitem{ZinnJustin:2002ru}
J.~Zinn-Justin, {\em {Quantum field theory and critical phenomena}}, Int. Ser.
  Monogr. Phys. {\bf 113} (2002) 1--1054.

\bibitem{Dorigoni:2014hea}
D.~Dorigoni, {\em {An introduction to resurgence, trans-series and alien
  calculus}}, {\tt arXiv:1411.3585} {\tt [hep-th]}.

\bibitem{Marino:2007te}
M.~Marino, R.~Schiappa, and M.~Weiss, {\em {Nonperturbative Effects and the
  Large-Order Behavior of Matrix Models and Topological Strings}}, Commun. Num.
  Theor. Phys. {\bf 2} (2008) 349--419, {\tt arXiv:0711.1954} {\tt [hep-th]}.

\bibitem{ZJJ1}
J.~Zinn-Justin and U.~D. Jentschura, {\em {Multi-instantons and exact results
  I: Conjectures, WKB expansions, and instanton interactions}}, Annals Phys.
  {\bf 313} (2004) 197--267, {\tt arXiv:quant-ph/0501136} {\tt [quant-ph]}.

\bibitem{ZJJ2}
J.~Zinn-Justin and U.~D. Jentschura, {\em {Multi-instantons and exact results
  II: Specific cases, higher-order effects, and numerical calculations}},
  Annals Phys. {\bf 313} (2004) 269--325, {\tt arXiv:quant-ph/0501137} {\tt
  [quant-ph]}.

\bibitem{Aganagic:2003qj}
M.~Aganagic, R.~Dijkgraaf, A.~Klemm, M.~Mari\~{n}o, and C.~Vafa, {\em
  {Topological strings and integrable hierarchies}}, Commun. Math. Phys. {\bf
  261} (2006) 451--516, {\tt arXiv:hep-th/0312085} {\tt [hep-th]}.

\bibitem{Aganagic2012}
M.~Aganagic, M.~C.~N. Cheng, R.~Dijkgraaf, D.~Krefl, and C.~Vafa, {\em {Quantum
  geometry of refined topological strings}}, JHEP {\bf 11} (2012) 019, {\tt
  arXiv:1105.0630} {\tt [hep-th]}.

\bibitem{Kallen2013}
J.~Kallen and M.~Mari\~{n}o, {\em {Instanton effects and quantum spectral
  curves}}, Annales Henri Poincare {\bf 17} (2016) 1037--1074, {\tt
  arXiv:1308.6485} {\tt [hep-th]}.

\bibitem{Kashaev:2015kha}
R.~Kashaev and M.~Mari\~{n}o, {\em {Operators from mirror curves and the
  quantum dilogarithm}}, {\tt arXiv:1501.01014} {\tt [hep-th]}.

\bibitem{Wang2015}
X.~Wang, G.~Zhang, and M.-x. Huang, {\em {New exact quantization condition for
  toric Calabi-Yau geometries}}, Phys. Rev. Lett. {\bf 115} (2015) 121601, {\tt
  arXiv:1505.05360} {\tt [hep-th]}.

\bibitem{Sun2016}
K.~Sun, X.~Wang, and M.-x. Huang, {\em {Exact quantization conditions, toric
  Calabi-Yau and nonperturbative topological string}}, {\tt arXiv:1606.07330}
  {\tt [hep-th]}.

\bibitem{Grassi:2016nnt}
A.~Grassi and J.~Gu, {\em {BPS relations from spectral problems and blowup
  equations}}, {\tt arXiv:1609.05914} {\tt [hep-th]}.

\bibitem{Grassi:2017qee}
A.~Grassi and M.~Mari\~no, {\em {The complex side of the TS/ST
  correspondence}}, {\tt arXiv:1708.08642} {\tt [hep-th]}.

\bibitem{Marino:2017gyg}
M.~Mari\~{n}o and S.~Zakany, {\em {Wavefunctions, integrability, and open
  strings}}, {\tt arXiv:1706.07402} {\tt [hep-th]}.

\bibitem{Marino:2016rsq}
M.~Mari\~{n}o and S.~Zakany, {\em {Exact eigenfunctions and the open
  topological string}}, J. Phys. {\bf A50} (2017) 325401, {\tt
  arXiv:1606.05297} {\tt [hep-th]}.

\bibitem{Zakany:2017txl}
S.~Zakany, {\em {Quantized mirror curves and resummed WKB}}, {\tt
  arXiv:1711.01099} {\tt [hep-th]}.

\bibitem{Kashani-Poor:2016edc}
A.-K. Kashani-Poor, {\em {Quantization condition from exact WKB for difference
  equations}}, JHEP {\bf 06} (2016) 180, {\tt arXiv:1604.01690} {\tt [hep-th]}.

\bibitem{Sciarappa:2016ctj}
A.~Sciarappa, {\em {Bethe/Gauge correspondence in odd dimension: modular
  double, non-perturbative corrections and open topological strings}}, JHEP
  {\bf 10} (2016) 014, {\tt arXiv:1606.01000} {\tt [hep-th]}.

\bibitem{Marino:2015ixa}
M.~Mari\~{n}o and S.~Zakany, {\em {Matrix models from operators and topological
  strings}}, Annales Henri Poincare {\bf 17} (2016) 1075--1108, {\tt
  arXiv:1502.02958} {\tt [hep-th]}.

\bibitem{Kashaev:2015wia}
R.~Kashaev, M.~Mari\~{n}o, and S.~Zakany, {\em {Matrix models from operators
  and topological strings, 2}}, {\tt arXiv:1505.02243} {\tt [hep-th]}.

\bibitem{Marino:2015nla}
M.~Mari\~{n}o, {\em {Spectral theory and mirror symmetry}}, {\tt
  arXiv:1506.07757} {\tt [math-ph]}.

\bibitem{Codesido2015a}
S.~Codesido, A.~Grassi, and M.~Mari\~{n}o, {\em {Spectral theory and mirror
  curves of higher genus}}, {\tt arXiv:1507.02096} {\tt [hep-th]}.

\bibitem{Bonelli:2016idi}
G.~Bonelli, A.~Grassi, and A.~Tanzini, {\em {Seiberg-Witten theory as a Fermi
  gas}}, Lett. Math. Phys. {\bf 107} (2017) 1--30, {\tt arXiv:1603.01174} {\tt
  [hep-th]}.

\bibitem{Bonelli:2017ptp}
G.~Bonelli, A.~Grassi, and A.~Tanzini, {\em {New results in $\mathcal{N}=2$
  theories from non-perturbative string}}, {\tt arXiv:1704.01517} {\tt
  [hep-th]}.

\bibitem{Bonelli:2017gdk}
G.~Bonelli, A.~Grassi, and A.~Tanzini, {\em {Quantum curves and $q$-deformed
  Painlev\'e equations}}, {\tt arXiv:1710.11603} {\tt [hep-th]}.

\bibitem{Grassi:2018bci}
A.~Grassi and M.~Mari\~no, {\em {A solvable deformation of quantum mechanics}},
  {\tt arXiv:1806.01407} {\tt [hep-th]}.

\bibitem{nekrasov2010quantization}
N.~A. Nekrasov and S.~L. Shatashvili, {\em Quantization of integrable systems
  and four dimensional gauge theories}, XVIth International Congress On
  Mathematical Physics: (With DVD-ROM), World Scientific, 2010, pp.~265--289.

\bibitem{alvarez2002-quartic}
G.~\'{A}lvarez, C.~J. Howls, and H.~J. Silverstone, {\em {Dispersive
  hyperasymptotics and the anharmonic oscillator}}, Journal of Physics A:
  Mathematical and General {\bf 35} (2002) 4017.

\bibitem{alvarez2004-doublewell}
G.~{\'A}lvarez, {\em {Langer--Cherry derivation of the multi-instanton
  expansion for the symmetric double well}}, Journal of mathematical physics
  {\bf 45} (2004) 3095--3108.

\bibitem{alvarez2000-cubic}
G.~{\'A}lvarez and C.~Casares, {\em {Exponentially small corrections in the
  asymptotic expansion of the eigenvalues of the cubic anharmonic oscillator}},
  Journal of Physics A: Mathematical and General {\bf 33} (2000) 5171.

\bibitem{alvarez2000-generic}
G.~{\'A}lvarez and C.~Casares, {\em {Uniform asymptotic and JWKB expansions for
  anharmonic oscillators}}, Journal of Physics A: Mathematical and General {\bf
  33} (2000) 2499.

\bibitem{alvarez2002anharmonic}
G.~{\'A}lvarez, C.~J. Howls, and H.~J. Silverstone, {\em {Anharmonic oscillator
  discontinuity formulae up to second-exponentially-small order}}, Journal of
  Physics A: Mathematical and General {\bf 35} (2002) 4003.

\bibitem{Dunne2014}
G.~V. Dunne and M.~\"{U}nsal, {\em {Uniform WKB, multi-instantons, and
  resurgent trans-series}}, Phys. Rev. {\bf D89} (2014) 105009, {\tt
  arXiv:1401.5202} {\tt [hep-th]}.

\bibitem{coleman1988aspects}
S.~Coleman, {\em Aspects of symmetry: selected erice lectures}, Cambridge
  University Press, 1988.

\bibitem{Dunne:2007rt}
G.~V. Dunne, {\em {Functional determinants in quantum field theory}}, J. Phys.
  {\bf A41} (2008) 304006, {\tt arXiv:0711.1178} {\tt [hep-th]}.

\bibitem{Takhtajan:2008}
L.~A. Takhtajan, {\em {Quantum Mechanics for Mathematicians}}, American
  Mathematical Society, 2008.

\bibitem{Marino:2015yie}
M.~Mari\~no, {\em {Instantons and Large N}}, Cambridge University Press, 2015.

\end{thebibliography}

\end{document}